\documentclass[useAMS,usenatbib]{mn2e}

\usepackage{natbib}
\usepackage{amssymb,amsmath}
\usepackage{graphicx}
\usepackage{enumerate}

\providecommand{\nat}[0]{Nature}

\providecommand{\apj}[0]{Astrophys. J.}
\providecommand{\apjl}[0]{Astrophys. J. Lett.}
\providecommand{\apjs}[0]{Astrophys. J. Supp. Ser. }

\providecommand{\aap}[0]{Astron. Astrophys. }

\providecommand{\physrep}[0]{Phys. Rep. }
\providecommand{\mnras}[0]{Mon. Not. Roy. Astron. Soc. }

\providecommand{\prl}[0]{Phys. Rev. Lett.}
\providecommand{\prd}{Phys. Rev. D.}

\providecommand{\physrep}[0]{Phys. Rep.}
\providecommand{\ssr}[0]{Space Sci. Rev.}

\def\beq{\begin{equation}}
\def\enq{\end{equation}}
\def\bea{\begin{eqnarray}}
\def\ena{\end{eqnarray}}

\def\L54{L_{54}}
\def\E55{E_{55}}
\def\et3{\eta_3}
\def\th1{\theta_{-1}}
\def\r07{r_{0,7}}
\def\x05{x_{0.5}}
\def\et600{\eta_{600}}
\def\et3{\eta_3}

\def\Fl{\mathcal{F}}

\def\cm{\hbox{~cm}}

\title[CTA is Well Suited to Follow Up Gravitational Wave Transients]{Cherenkov Telescope Array is Well Suited to Follow Up Gravitational Wave Transients}

\author[I. Bartos et al.]{I. Bartos$^{1,2}$\thanks{ibartos@phys.columbia.edu}, P. Veres$^3$, D. Nieto$^2$, V. Connaughton$^4$, B. Humensky$^2$, K. Hurley$^5$, \newauthor
S. M\'{a}rka$^{1,2}$, P. M\'{e}sz\'{a}ros$^3$, R. Mukherjee$^6$, P. O'Brien$^7$, J.P. Osborne$^7$\\
$^1$Department of Physics, Columbia University, New York, NY 10027, USA \\
$^2$Columbia Astrophysics Laboratory, Columbia University, New York, NY 10027, USA \\
$^3$Department of Astronomy and Astrophysics, Pennsylvania State University, University Park, PA 16802, USA \\
$^4$Center for Space Plasma and Aeronomic Research (CSPAR), University of Alabama in Huntsville, Huntsville, AL 35899, USA \\
$^5$University of California-Berkeley, Space Sciences Laboratory, Berkeley, CA 94720, USA \\
$^6$Department of Physics and Astronomy, Barnard College, Columbia University, New York, NY 10027, USA \\
$^7$Space Research Centre, Department of Physics \& Astronomy, University of Leicester, Leicester LE1 7RH, UK}

\begin{document}

\date{}

\pagerange{\pageref{firstpage}--\pageref{lastpage}} \pubyear{2002}

\maketitle

\label{firstpage}

\begin{abstract}
The first gravitational-wave (GW) observations will greatly benefit from the detection of coincident electromagnetic counterparts. Electromagnetic follow-ups will nevertheless be challenging for GWs with poorly reconstructed directions. GW source localization can be inefficient (i) if only two GW observatories are in operation; (ii) if the detectors' sensitivities are highly non-uniform; (iii) for events near the detectors' horizon distance. For these events, follow-up observations will need to cover 100-1000 square degrees of the sky over a limited period of time, reducing the list of suitable telescopes. We demonstrate that the Cherenkov Telescope Array will be capable of following up GW event candidates over the required large sky area with sufficient sensitivity to detect short gamma-ray bursts, which are thought to originate from compact binary mergers, out to the horizon distance of advanced LIGO/Virgo. CTA can therefore be invaluable starting with the first multimessenger detections, even with poorly reconstructed GW source directions. This scenario also provides a further scientific incentive for GW observatories to further decrease the delay of their event reconstruction.
\end{abstract}


\section{Introduction}

Observing the electromagnetic counterparts of the first detected gravitational-wave (GW) signals is one of the major goals of astronomy for the near future \citep{2009arXiv0902.1527B,2012ApJ...759...22K}. Electromagnetic counterparts would greatly increase our confidence in the first detection, and could revolutionize our understanding of some cosmic phenomena (e.g., \citealt{2012A&A...541A.155A,2012ApJS..203...28E,2013CQGra..30l3001B,2013arXiv1310.2314T}).

One of the most anticipated cosmic phenomena that is expected to result in the first GW detections is the merger of two compact stellar-mass objects, which are either neutron stars or black holes \citep{2012PhRvD..85h2002A}. Of special interest are binaries that consist of at least one neutron star, which can also produce electromagnetic radiation \citep{1984SvAL...10..177B,1986ApJ...308L..43P,1989Natur.340..126E,2007NJPh....9...17L,2013CQGra..30l3001B} as well as other messengers, such as cosmic rays or neutrinos \citep{1997PhRvL..78.2292W,PhysRevLett.108.231101,2012ApJ...752...29H,PhysRevLett.107.251101,2012arXiv1203.5192A,2013PhRvL.111m1102M}.

Several promising emission processes have been suggested that would accompany compact binary mergers. First of all, short gamma-ray bursts (GRBs; \citealt{2013APh....43..134M}) are thought to originate from these mergers. Gamma rays can be produced in the outflows driven by an accreting black hole that forms in the merger (e.g., \citealt{2007PhR...442..166N}). The afterglows of some of these short GRBs, produced by the interaction of the outflow with the ambient medium, represent an additional electromagnetic counterpart \citep{1999ApJ...520..641S,2011ApJ...733L..37V}. Further, energetic, sub or mildly relativistic outflows launched by a binary merger can also interact with the surrounding medium, producing quasi-isotropic emission in the radio band over a period of more than a year \citep{2011Natur.478...82N,2013MNRAS.430.2121P}. The same outflow can also undergo r-process nucleosynthesis during its expansion, resulting in near-infrared--infrared radiation, called a kilonova (also called macronova; \citealt{1998ApJ...507L..59L,2005astro.ph.10256K,2008MNRAS.390..781M,2010MNRAS.406.2650M,2013ApJ...774...25K,2013MNRAS.435..502F,tanvir2013kilonova,2013arXiv1311.2603B}).

With the available observational capabilities, a large fraction of the produced electromagnetic counterparts may only be detectable, if detectable at all, with follow-up observations guided by GW triggers \citep{2008CQGra..25r4034K,2011ApJ...733L..37V,2012ApJ...746...48M,2012A&A...539A.124L,2012ApJS..203...28E,2013CQGra..30l3001B}. The direction reconstruction of GW detectors, however, is limited to $\gg1$\,deg$^2$ \citep{2013arXiv1304.0670L}. This substantially reduces the feasibility of many electromagnetic follow-up efforts, given the limited field of view of the most sensitive telescopes, and the limited sensitivity of larger-field-of-view telescopes. Nevertheless, a number of telescopes may be competitive at following up GW triggers with very low latency with strategies optimized to cover a significant fraction of the GW sky area. These include moderate-aperture telescopes with large field of view such as the BlackGEM Array\footnote{https://www.astro.ru.nl/wiki/research/blackgemarray} and the Ground Wide Angle Cameras (GWAC; \citealt{2009AIPC.1133...25G}) that will be dedicated follow-up operations. Highly sensitive instruments with limited field of view, such as Swift, may also be promising follow-up facilities \citep{2012ApJS..203...28E}. Another interesting direction is the Ultra Fast Flash Observatory \citep{2012SPIE.8443E..0IP}, which will have a large field-of-view x-ray detector as well as sub-second optical follow-up capability.

Following up the first GW observations, probably around 2016-2018 \citep{2013arXiv1304.0670L}, will be particularly challenging. At this time, given the staged schedule of construction and commissioning of GW detectors, it is possible that direction reconstruction of the first detections will mainly rely on a two-detector network (or a three-detector network with highly non-uniform sensitivity; \citealt{2013arXiv1304.0670L}). This will substantially decrease the location accuracy of GW measurements, necessitating electromagnetic follow-ups with large, $100-1000$\,deg$^2$, search areas at high sensitivity. Further, the direction of a GW event will typically be localized to multiple, disjoint sky regions at potentially distant parts of the sky, requiring follow-up observations to cover these separate sky regions. Radio follow-up observations, e.g., with the Square Kilometre Array (SKA; \citealt{2003ASPC..289...21E}) or LOFAR \citep{2009IEEEP..97.1431D}, are another interesting alternative, given the expected long-lived radio emission following the binary merger \cite{2013MNRAS.430.2121P}.

In this paper we propose and investigate the possibility for large-field-of-view electromagnetic follow-up observations of GW event candidates, using the Cherenkov Telescope Array (CTA; \citealt{2011ExA....32..193A}). CTA is well suited for GW-follow-up observations for multiple reasons:
\begin{enumerate}
\item \emph{Field-of-view:} CTA will be capable of monitoring a large sky area via survey mode operation (either by pointing its telescopes in different directions, or by rapidly scanning a set of consecutive directions). It will be able to monitor the $\sim$$1000$\,deg$^2$ area necessary for early GW triggers for which only an incomplete GW detector network is available. This survey mode will also be useful for later GW observations: since localization becomes less efficient with, e.g., decreasing signal-to-noise ratio (SNR), a significant fraction of GW event candidates will have large error regions even when more than two GW detectors are available.
\item \emph{Coincident observational schedule:} CTA is expected to begin partial operation around 2017, therefore it will probably be available to follow up the first GW detections. The anticipated completion of CTA is around 2020.
\item \emph{Rapid response:} CTA has the capability to respond to target-of-opportunity requests and start monitoring the selected sky area within $\sim$\,$30$\,sec \citep{2013APh....43..317D}. This is important given the limited duration ($\lesssim1000$\,s; Section \ref{section:CTAsensitivity}) of high-energy photon emission connected with GRBs. The sensitivity of CTA to GRBs will be mainly determined by its so-called Large Size Telescopes (LST; \citealt{2013APh....43....3A}), which are capable of the fastest response ($180^\circ$ slewing in less than 20\,s; \citealt{Inoue+13GRBCTA}).
\end{enumerate}
In the following we discuss these points further in detail.

The use of CTA to follow up GW event candidates has been previously suggested by \cite{2013APh....43..189D}.
In this paper, we explore in detail the follow-up of GW event candidates by CTA, and the particular advantage of CTA in following-up poorly localized signals.

The paper is organized as follows. In Section \ref{section:GW} we discuss the expected sensitivity and localization capability of advanced GW detectors. In Section \ref{section:CTAsensitivity} we estimate the sensitivity of CTA for detecting short GRBs with known directions, exploring multiple emission models focusing on the distances relevant for GW detection. In Section \ref{section:CTAsurvey} we describe the sensitivity of CTA in survey mode, focusing on directional uncertainties relevant for GW searches. Section \ref{section:GBM} discusses the role of satellite-based GRB detectors in adding information to the follow-up process. Finally, Section \ref{section:conclusion} summarizes our results and presents our conclusions.

\section{Gravitational-wave Detection from Compact Binary Coalescences}
\label{section:GW}
\subsection{Sensitivity}

In order to understand the characteristic distances at which the first GW detections are anticipated, we briefly review the expected sensitivity of advanced GW detectors. This sensitivity is expected to be relatively low at the start of operation in 2015, and will gradually increase to design sensitivity towards 2019 \citep{2013arXiv1304.0670L}.

For estimating detection sensitivities, in the following we focus on neutron star binary mergers. We will conservatively assume that all short GRBs originate from neutron star binary mergers, noting that the GW horizon distance for black-hole--neutron-star mergers is greater. To characterize the sensitivity of a GW detector we use the so-called horizon distance: the distance to which a source at optimal location and with optimal orientation is detectable with single-detector SNR 8 (e.g., \citealt{2010CQGra..27q3001A}). This SNR corresponds to a false alarm rate of $\ll1$\,yr$^{-1}$ for a network of 2 detectors \citep{2013arXiv1304.0670L}, making it a useful limit for follow-up observations. We note here that direction-averaged sensitivity of a GW detector is $\sim$$1.5$ times less than the horizon distance, which we take into account in detection rate estimates below. We assume that the rotation axes of the mergers approximately point towards the earth, given that short GRBs are expected to be beamed. We consider a short-GRB rate of $\gtrsim10$\,Gpc$^{-3}$yr$^{-1}$ for bursts that are beamed towards the earth \citep{2006ApJ...650..281N}. Below we outline the GW observation schedule and detection prospects of neutron star binary mergers, based on the schedule presented by \cite{2013arXiv1304.0670L}.

2015: the horizon distance of Advanced LIGO (aLIGO) detectors during the first observation period in 2015 is expected to be limited: 90 -- 180\,Mpc. It is highly unlikely that a short GRB will occur within the corresponding detection volume of $10^{-4}$--$10^{-3}$\,Gpc$^{3}$ during the expected 3 months of observation. 

2016-2017: 6 months of joint aLIGO-Virgo observation run are envisioned, with aLIGO horizon distance of 180--270\,Mpc, and a few times smaller horizon distance for Advanced Virgo. This corresponds to a detection volume of $0.01$--$0.02$\,Gpc$^3$. Within this volume there may be a binary neutron star merger during the observation time. A short GRB beamed towards the earth within this volume is also possible (with probability $p \gtrsim 5-10\%$).

2017-2018: a $\sim$$9$-months long GW observation period will take place. The horizon distance of aLIGO will be 270--380\,Mpc, twice as much as the sensitivity of Virgo. There will likely be multiple neutron star binary mergers within the corresponding detection volume of 0.03--0.1\,Gpc$^3$, with a good chance (p$\gtrsim20$--$50\%$) of a short GRB beamed towards the earth within this volume and observation time.

2019+: extended observation at design sensitivity, with horizon distance $\sim$$450$\,Mpc. This corresponds to a detection volume of $\gtrsim0.1$\,Gpc$^{3}$. On average, more than one short GRB beamed towards the Earth is expected within this volume for every year of operation. This sensitivity will likely further increase with the completion of additional GW detectors KAGRA \citep{2011IJMPD..20.1755K} and aLIGO India \citep{LVCcommissioning}.

\subsection{Localization}

The waveform of a GW signal detected by individual GW detectors is greatly degenerate with respect to source direction. To recover the direction of a GW event, multiple GW observatories are used, mainly taking advantage of the different time-of-arrival of the GW signal at different detectors (\citealt{2013arXiv1304.0670L} and references therein).

In the literature, localization is mainly discussed for the case of three or more GW detectors with similar sensitivities \citep{2006PhRvD..74h2004C,2007PhRvD..75f2004R,2011PhRvD..83j2001K,2011ApJ...739...99N,2011PhRvD..84j4020V,2011CQGra..28l5023S,2012PhRvD..85j4045V,2013ApJ...767..124N}. For the three-detector case, localization of a signal with single-detector SNR\,$=$\,8 is typically $\sim$$100$\,deg$^{2}$ ($90\%$ confidence; \citealt{2011CQGra..28j5021F}). A fraction ($\sim$10$\%$) of the signals can be localized with higher precision ($\sim$$10$\,deg$^{2}$; \citealt{2013arXiv1304.0670L,2013ApJ...767..124N}). For black-hole--neutron-star binaries, the black-hole spin can further improve the precision of localization \citep{2008ApJ...688L..61V,2009CQGra..26k4007R}; spin, however, does not affect localization for neutron star binaries.

For cases in which the reconstructed sky region is too large for feasible follow-up, a possible strategy is to decrease the sky region by increasing the false dismissal rate of the observation\footnote{I.e. focusing on a smaller sky region, the probability that the real source direction is within this sky region becomes smaller.}. In practice, the size of the reconstructed sky region can be significantly reduced if one focuses on encompassing a lower integrated confidence region. As an example \cite{2011PhRvD..83j2001K}, studying the localization of GW transients with 3 and 4-detector networks, found that the sky area corresponding to localization with $50\%$ confidence can be significantly smaller (up to $\times$10) than the sky area encompassing $90\%$ confidence level.

For the first observation runs with advanced GW detectors, as well as for some of the observation time later on, one can expect to have an essentially two-detector GW detector network. This will be the case early on due to the different construction schedules of aLIGO and Virgo, and partially later on due to the sub-$100\%$ duty cycle of each individual observatory. It is therefore important to explore the localization capability of GW event candidates with using only two GW detectors.

Two-detector observations of GW sources, by only applying timing constraints, can constrain source direction to essentially a ring on the sky  \citep{2008ApJ...688L..61V,2009NJPh...11l3006F,2009CQGra..26t4010V,2010PhRvD..81h2001W,2011CQGra..28j5021F}. It is possible to further constrain the sky area by using amplitude and phase information \citep{2009CQGra..26t4010V,2013arXiv1309.1554K}, but for practical purposes this will still leave a large localization uncertainty \citep{2013arXiv1304.0670L,2013arXiv1309.1554K} that is difficult to cover for many electromagnetic follow-up facilities. A recent numerical study by \cite{2013arXiv1309.1554K}, applying idealized Gaussian background noise and utilizing amplitude and phase information, shows that, using the two aLIGO detectors only, neutron star binary mergers with network-SNR$>12$ can be localized to within $100-1000$\,deg$^{2}$ (with 95$\%$ confidence), with a median localization of $250$\,deg$^{2}$.

A further complication is that the reconstructed sky area is typically discontinuous, due to the direction-dependent detector sensitivity. Parts of the sky area can be distributed over a large range of angles, raising an additional challenge to follow-up facilities.

Based on the results of \cite{2011CQGra..28j5021F}, we estimate the typical localization sky area for a two-detector network (LIGO Hanford and LIGO Livingstone) to be $\sim$$2000$\,deg$^2$ for single-detector SNR$= 8$ at $90\%$ confidence level (\citealt{2011CQGra..28j5021F} considered two detectors at Hanford and one at Livingston; we converted this result to the two-detector-only case, which is $\sim$$\sqrt{3/2}$ greater). This sky area $\Omega$ scales with the SNR (see \citealt{2010PhRvD..81h2001W}) and with confidence level CL (see \citealt{2011CQGra..28j5021F}) as
\begin{equation}
\Omega\approx 2000\,\mbox{deg}^2 \left(\frac{8}{\mbox{SNR}}\right) \left(\frac{\mbox{erf}^{-1}(\mbox{CL})}{\mbox{erf}^{-1}(0.9)}\right),
\end{equation}
where $\mbox{erf}^{-1}$ is the inverse error function.

For sources within $\sim$$100$\,Mpc, another possible way of decreasing the localization uncertainty could be to focus on galaxy locations within the error region of the GW signal (e.g., \citealt{2008CQGra..25r4034K}). While this can decrease the set of directions that follow-up observatories need to survey, the benefit from focusing on galaxies decreases for larger distances and sky areas due to the following
reasons: (i) for larger distances the area-density of galaxies becomes comparable to the field of view of follow-up telescopes, therefore there is no added benefit in looking at them individually; (ii) galaxy catalogs are incomplete beyond a few tens of megaparsecs \citep{2011CQGra..28h5016W}, therefore the hosts of some GW events will be overlooked by a galaxy-based follow-up search; (iii) with a GW sky area of $\sim$$1000$\,deg$^{2}$, the number of galaxies within $200$\,Mpc is $\mathcal{O}(10^6)$ \citep{2013ApJ...767..124N}, making galaxy-based searches on this scale impractical for large sky areas\footnote{Note that, for three or more available detectors and sufficiently high SNR, which is the likely case for some sources after $\sim2020$, galaxies will be useful even at the $200$\,Mpc scale \citep{2013ApJ...767..124N}.}.

\section{Detecting Short Gamma-ray Bursts with CTA}
\label{section:CTAsensitivity}

So far multi-GeV electromagnetic emission has been detected only from a fraction of the short GRBs observed in the keV-MeV energy band (e.g., 081024B and 090510; \citealt{2010ApJ...712..558A,2010ApJ...716.1178A,2011ApJ...730..141Z}). While this could be a consequence of their intrinsic emission mechanism, it could also be a result of the limited sensitivity of current high-energy gamma-ray observatories. Due to the limitations of available observations, in this section we calculate the multi-GeV photon emission from short GRBs based on their observed lower-energy emission. Based on these calculations, we then estimate their detectability using the Cherenkov Telescope Array (CTA; see ~\citealt{Acharya20133} and
references therein). We first consider the sensitivity of CTA when it is pointing at one particular direction. We then additionally take into account the effect of having a poorly reconstructed direction, and the change in sensitivity when the detector surveys an extended sky area.

\subsection{The Cherenkov Telescope Array}

CTA is an international project leading to the
realization of a new observatory for very high energy (VHE) gamma
rays. CTA represents the next generation of imaging atmospheric
Cherenkov telescopes (IACTs; for a historical review
see~\citealt{Hillas201319}), providing an order of magnitude improvement in sensitivity over the
current-generation IACTs, along with improved
angular and energy resolutions, and spanning about four decades in energy
(from a few tens of GeV to above 100\,TeV).
CTA relies on the technique of imaging the Cherenkov light flashes
emitted by the particle showers induced in the atmosphere by impinging
gamma-rays, reconstructing the primary gamma-ray's energy and arrival
direction from several such images formed in the camera plane.

CTA plans to operate two sites, one in the northern hemisphere and one in the southern hemisphere, which together will provide
full-sky coverage. Each installation will consist of an array of
50-100 telescopes in three different sizes, each one optimized for a
particular energy range. CTA has begun its prototyping phase and the
construction phase is expected to start in early 2015 with an
estimated completion date of 2020, although the scientific studies may
start as early as 2016.

It is worth noting that, for the first time in the field of
ground-based VHE gamma-ray instruments, CTA will operate as an open
observatory, allowing the entire scientific community to request
observation time, as well as granting public access to the CTA data


\subsection{Observed Multi-GeV Photon Emission from Short GRBs}


The Large Area Telescope on the Fermi satellite (hereafter Fermi-LAT; \citealt{2009ApJ...697.1071A}) has detected very high-energy emission from six short GRBs (out of $\sim70$ GRBs detected at keV-MeV energies by the Fermi Gamma-ray Burst Monitor, hereafter Fermi-GBM; see public table\footnote{{http://fermi.gsfc.nasa.gov/}}).  Two of these were detected in the less stringent class of events, the LAT Low Energy (LLE) data,  with the other four detected in a standard Transient class analysis above 100\,MeV.   Two were seen above 1\,GeV.

High-energy emission from short GRBs appears to be of two varieties: prompt emission that is coincident with the peaks seen in the Fermi-GBM data (though the first peak is often missing at high energies) and emission that persists for up to 100\,s \citep{2010ApJ...716.1178A}, decaying smoothly over time.  Since Fermi-LAT is fluence limited, it is less likely to detect short GRBs, and thus the fraction of short-to-long GRBs seen in Fermi-LAT is much smaller ($< 10\%$) than that seen at lower, $\sim$\,MeV energies by Fermi-GBM ($18\%$).   The fraction of high-energy emission ($\gtrsim 100$\,MeV) to kev-MeV emission (in the band [10\,keV, 1\,MeV]), however, is higher for short GRBs than for long GRBs (well above $10\%$, sometimes above $100\%$ compared to $10\%$ for long GRBs). This is mostly because the high-energy emission is extended in time whereas the prompt emission is contained within a 2\,s period (see \citealt{2010ApJS..188..405A}).  Thus there appears to be a threshold effect for short GRBs based on the sensitivity of Fermi-LAT to the short-lived prompt emission, but those short GRBs that rise above the threshold exhibit interesting behavior that makes them promising candidates for a more sensitive instrument, particularly in their long-lived emission.

Although of the short GRBs only GRB\,090510 can be studied in detail, owing to poor the photon statistics for the other cases, the extended emission in LAT-detected GRBs in general appears to be spectrally harder than the prompt emission, with multi-GeV photons often detected hundreds of seconds after the prompt emission (e.g., \citealt{Ackermann03012014,0067-0049-209-1-11}). LAT emission from the energetic GRB\,090510 ($E_{\rm kin}=10^{53}$\,erg) was observed for up to $\sim 100$ seconds \citep{2010ApJ...716.1178A,2011ApJ...730..141Z}, indicating that longer duration, high-energy emission is possible even for short GRBs. This is in contrast with the prompt MeV-range emission of short GRBs, which typically lasts for less than one second.

The decay of the GeV light curves of GRBs (long and short) generally follows a power law in time, of index typically between -1.1 and -1.4. This behavior is the same as expected from the decay of an external shock afterglow (e.g., \citealt{2010MNRAS.409..226K,2010MNRAS.403..926G,2012RAA....12.1139M}). The typical photon spectral indices above the ($\sim$\,MeV) spectral peak are in the range of -2.1 to -2.6, and in only one case has a high energy spectral steepening been detected in the GeV range (in GRB 090926A; \citealt{2011ApJ...729..114A}). It is unclear whether this is due to source-intrinsic effects or to external absorption; there is no evidence for a spectral turnover or energy cutoff in any other Fermi-LAT GRB observed so far.

\subsection{Possible Origin and Properties of Multi-GeV Photon Emission}

Only the first few seconds of the observed LAT emission could originate from a collisional photosphere \citep{2010MNRAS.407.1033B}, since the outflow duration is generally associated with the duration of the MeV prompt emission. Leptonic GeV emission by upscattering of photospheric soft photons by internal shocks
\citep{2011MNRAS.415.1663T} or external shocks \citep{2013ApJ...764...94V} could last longer than the
photospheric emission, but at most by the angular time corresponding to the shock radius,
$r_{shock}/c\Gamma^2$~($\Gamma$ is the Lorentz factor of the outflow), i.e., a few seconds to $\lesssim 10$\,s -- after that, the forward
shock synchrotron and synchrotron-self Compton emission would take over the GeV emission.
In an external shock the photon-photon self absorption generally sets in at observer-frame
photon energies $\gtrsim$\,TeV \citep{2001ApJ...559..110Z}; this is dependent on the bulk Lorentz
factor (e.g., \citealt{2010ApJ...716.1178A,2012MNRAS.421..525H}), and it is additional
to any external (induced by extragalactic background light, hereafter EBL) absorption, which also sets in at similar energies.

Searches for TeV emission associated with GRBs so far have yielded only upper limits
(e.g., \citealt{Albert+06magic050713a,Abdo+12milagro,2011ApJ...743...62A}). The Fermi-LAT observation of
0.1\,TeV photons from the nearby (z=0.3) GRB130427a shows
that GRB spectra can extend up to $\sim$TeV energies, and GRBs are possible
sources in this energy domain (e.g., \citealt{Inoue+13GRBCTA}).

Since CTA observations are expected to follow a GW trigger with a delay of $\sim 100$\,s, we focus on the afterglow emission of GRBs, which also contains multi-GeV photons. The
prompt emission observations can be facilitated by suitable precursor
emissions, not uncommon in some cases of short GRBs \citep{Troja+10precursor}.

According to EBL models \citep{2011MNRAS.410.2556D,Stecker+06ebl}, TeV photons at $\sim$$100$\,Mpc will not
be significantly affected by pair annihilation with the EBL, while TeV sources
from $\sim$$400$\,Mpc will have significant annihilation.  Photons at sub-TeV energies ($\sim$$0.1$\,TeV), coming from the largest sensitivity range of aLIGO/Virgo, will not be affected by EBL.

Since there has been no confirmed TeV-photon detection from GRBs so far, the spectral properties of GRBs in the TeV domain are largely uncertain.  Nevertheless, for both short and long GRBs, there are cases in which a power-law component with a spectrum harder than $\gamma=-2, dN/dE \propto E^{\gamma}$ shows no cutoff up to multi-GeV energies \citep{Abdo+09-090902B}.  This is encouraging for future CTA observations.

\subsection{Observation Latency}

GeV emission is expected to start shortly after the merger of the compact binary, and it gradually fades away. CTA can therefore detect the most GeV photons if it commences the follow-up observation as soon as possible after the detection of the GW signal of the binary.

GW triggers for electromagnetic follow-up observations by initial LIGO/Virgo were distributed to observatories only $\gtrsim10$\,min after detection \citep{2012A&A...541A.155A,2012ApJS..203...28E}. The delay was mostly due to human monitoring.

An important ongoing effort for advanced GW analyses is the reduction of the latency of GW triggers distributed to follow-up observatories. Motivated by the electromagnetic emission following short GRBs shortly after the prompt emission, direction reconstruction methods aims to identify the sky area of GWs on less than a minute time scales. For instance BAYESTAR \citep{2014arXiv1404.5623S}, a rapid direction reconstruction algorithm was demonstrated to produce accurate sky areas in less than a minute after the detected merger of a binary neutron star. BAYESTAR will be used to reconstruct source directions from the beginning of advanced GW observations.

In the future, the delay from GW detectors can decrease even further \citep{2012ApJ...748..136C}. GW search algorithms will be capable of searching for the inspiral of a binary neutron star system even before its merger in practically real time. For sufficiently strong GW signals this can allow for the detection and parameter estimation of the binary, even before the actual merger of the neutron stars, resulting in even shorter triggering of electromagnetic follow-ups.

Given the above delay of $\sim 1$\,min for even early advanced GW observations, in the following we consider a conservative observational delay of $t_{\rm start}=100$\,s following the binary merger, associated with a 1-minute delay due to GW data analysis/detection and $\sim1/2\,$min delay due to the slewing of CTA. Nevertheless, we will also explore how changing $t_{\rm start}$ affects sensitivity.

\subsection{Sensitivity of CTA}

We estimate the sensitivity of CTA based on Fig.\,6 (\emph{upper}; best performance curve) of \cite{2013APh....43..171B}. This is the most recent public estimate of the sensitivity curve of CTA, produced by the CTA Monte Carlo working group using their baseline analysis method with the most up-to-date (public) Monte Carlo simulations of the array.

The GRB energy spectrum will likely cut off at some high-energy threshold (e.g., \citealt{2011ApJ...730..141Z}), therefore we consider three scenarios, in which the cutoff energies are $E_{\rm cutoff}=\{50, 100, 1000\}$\,GeV. Up to these thresholds, we assume that the GRB photon energy spectrum follows a power law (see above). Since the expected differential sensitivity improves with photon energy faster than $E^{2}$ \citep{2013APh....43..171B} in the energy range of interest, the greatest contribution to detection confidence will come from the highest energies just below the cutoff energy. We therefore conservatively approximate the sensitivity of the detector with its differential sensitivity in the highest energy bin below the threshold\footnote{Differential sensitivity is calculated by \cite{2013APh....43..171B} for energy intervals of equal log-scale width (0.2 in 10-based logscale). For cutoff energy $E_{\rm cutoff}$, we take the energy interval $\Delta E$ to be $\Delta E = (1-10^{-0.2}) E_{\rm cutoff}\approx 0.37 E_{\rm cutoff}$.}, noting that the integral sensitivity, using the complete energy range of the instrument, can be higher. For the $T_0=50$\,h exposure shown by \cite{2013APh....43..171B}, the expected differential sensitivity $\Phi_0$ corresponding to the three energy thresholds is $E^2\Phi_0= \{60,17,1.5\} \times 10^{-13}$\,erg\,cm$^{-2}$\,s$^{-1}$, respectively.
Following the method described in \cite{Gou+07glast}, we calculate the sensitivity for an exposure time $t_{\exp}$ by determining the minimum fluence $\Fl(t_{\exp})$ with which a source is detectable:
\begin{equation}
\Fl \approx \kappa \Phi_0 \left(\frac{T_0}{t_{\rm exp}}\right)^{1/2} E_{\rm cutoff}^2 t_{\rm exp}.
\end{equation}
where the constant $\kappa\leq 1$ depends on the width of the energy bin, as well as the spectral shape of the detector sensitivity and the photon flux from the source (it is $\leq1$ as long as the detector sensitivity decreases faster than the source spectrum). Below, we conservatively use $\kappa=1$. In the following we adopt a multi-GeV emission duration of $t_{\rm exp}=1000$\,s. While this is longer than the duration of extended emission for observed for most GRBs, this may be due to the limited sensitivity of Fermi-LAT.
With $t_{\rm exp}=1000$\,s, the limiting fluences for detection are:
\begin{equation}
\Fl_{\rm 1000s} \approx\{80,23, 2\}\times10^{-9}\,\mbox{erg\,cm}^{2}.
\label{eq:lim1k}
\end{equation}

\begin{figure}
  \resizebox{0.47\textwidth}{!}{\includegraphics{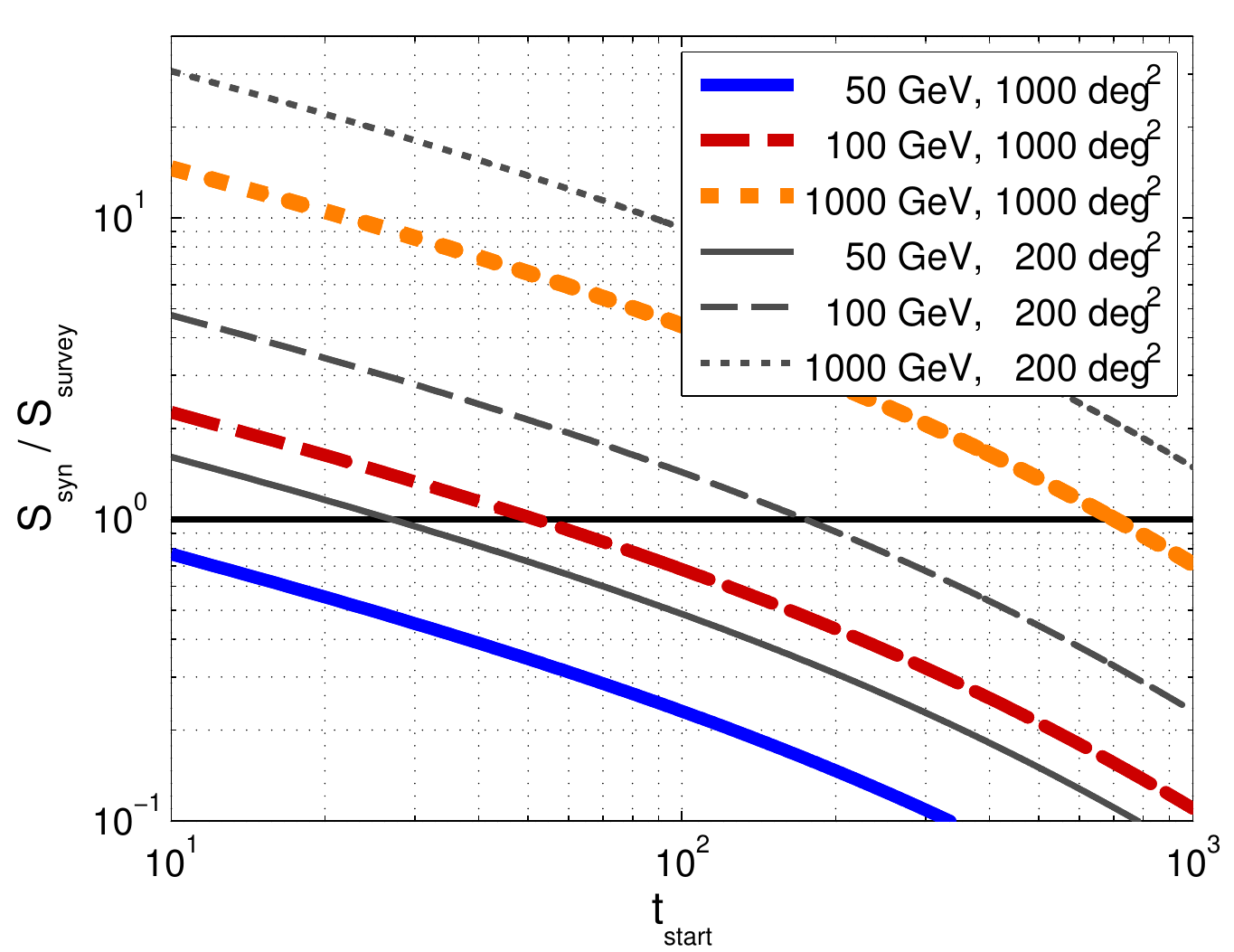}}
  \caption{Detectability of short GRBs with CTA as a function of the delay $t_{\rm start}$ between the onset of the GRB and the start of observations, for different high-energy emission cutoff energies ($E_{\rm cutoff}$). Results are shown for surveys over $\sim$$1000$\,deg$^2$ as well as $\sim$$200$\,deg$^2$ (see legend). The GRB ($E_{\rm kin}=10^{51}$\,erg; $D_{\rm L}=300$\,Mpc) is considered detectable if its fluence $S_{\rm syn}$ (estimated assuming synchrotron emission and cutoff energy $E_{\rm cutoff}$) is greater than the sensitivity $S_{\rm survey}$ of CTA in survey mode for an observation lasting for $1000$\,s. C.f. rows 5-7 in Table \ref{table:fluence}, where fluences are shown for a few characteristic $t_{\rm start}$ values.}\label{figure:thresholds}
\end{figure}

Above, the sensitivities are given for observations at $20^\circ$ zenith angle (in the frame of CTA). The zenith angle will affect the lower energy threshold at which CTA becomes sensitive (e.g., \citealt{2011arXiv1109.5680B}), therefore at higher zenith angles, the source may be detectable only if its emission extends to higher energies\footnote{\cite{2011arXiv1109.5680B} approximated the lower cutoff energy to depend on the zenith angle as $\propto \cos(\theta_{\rm zenith})^{-3}$.}. For simplicity, below we adopt the obtained sensitivities at $20^\circ$ for our analysis.

\begin{table*}
\centering
\begin{tabular}{l|ccc}
\hline\hline
$E_{\rm cutoff}$ (energy cutoff)                                                         & 50\,GeV 	 & 100\,GeV & 1\,TeV    \\
\hline\hline 
$S_{\rm det}$ (CTA)  [$10^{-9}$\,erg\,cm$^{-2}$]                                         & 80 		 & 23  	    & 2         \\
$S_{\rm survey}$ (CTA - survey; 1000 deg$^2$)                                            & 800		 & 230 		& 20        \\
$S_{\rm survey}$ (CTA - survey;  200 deg$^2$)                                            & 360		 & 100 		& 9         \\
\hline 
$S_{\rm syn}$ (GRB\,090510-like; $D_{\rm L}=5.8$\,Gpc)		                             & 50 		 & 40  		& 20        \\ 
$S_{\rm syn}$ (GRB\,090510-like; $D_{\rm L}=300$\,Mpc)  	                             & 33\,000 	 & 28\,000  & 16\,000   \\ 
$S_{\rm syn}$ ($E_{\rm kin}=10^{51}$\,erg; $D_{\rm L}=300$\,Mpc)  	                     & 190 		 & 160 	    & 90        \\ 
$S_{\rm syn}$ ($E_{\rm kin}=10^{51}$\,erg; $D_{\rm L}=300$\,Mpc; $t_{\rm start}=30$\,s)  & 360       & 310      & 170       \\ 
$S_{\rm syn}$ ($E_{\rm kin}=10^{51}$\,erg; $D_{\rm L}=300$\,Mpc; $t_{\rm start}=10^3$\,s)& 30 	     & 25 		& 14        \\
\hline 
$S_{\rm ssc}$ ($n=10^{-1}$\,cm$^{-3}$)								                     & 4400 	 & 5300	    & 3700      \\ 
$S_{\rm ssc}$ ($n=10^{-3}$\,cm$^{-3}$)					                                 & 20	 	 & 40 	    & 70        \\ 
\hline\hline
\end{tabular}
\caption{Fluence values for different emission and detection scenarios presented in the text, in units of $10^{-9}$\,erg\,cm$^{-2}$, for different emission cutoff energies $E_{\rm cutoff}$. CTA detectable fluences for single pointing and survey mode ($S_{\rm det}$, $S_{\rm survey}$) are shown for $1000$\,s duration, and for $1000$\,deg$^{2}$ as well as $200$\,deg$^{2}$ observable sky area in the case of the survey mode (see Section \ref{section:survey}). These detection fluences are calculated conservatively from the differential sensitivity of CTA (see text). For the fluence of synchrotron emission ($S_{\rm syn}$), we take the parameters of GRB\,090510, and an observation starting at $t_{\rm start}=100$\,s  after the binary merger and lasting for $t_{\rm duration}=1000$\,s, except for the parameters stated explicitly in the table. For the fluence of SSC emission ($S_{\rm ssc}$), we consider isotropic-equivalent GRB energy $E_{\rm kin}=10^{51}$\,erg, an observation with $t_{\rm start}=100$\,s and $t_{\rm duration}=1000$\,s, and circumburst number density $n$ stated in the table.}
\label{table:fluence}
\end{table*}

\subsubsection{CTA limits from synchrotron emission}

Thus far the prototypical short GRB with detected GeV-photon emission is GRB\,090510 at $z=0.9$
($D_{\rm L} = 5.8$\,Gpc), for which the highest energy photon detected
was $30$\,GeV \citep{Abdo+09-090510}.  With its duration of a few seconds, the
prompt emission cannot be realistically observed by CTA in follow-up mode (a serendipitous pointing in the right direction at the right time could, nevertheless, lead to detection).  The afterglow, however, is more promising. It follows
a power-law decay (temporal index: $\alpha = -1.38$; \citealt{2010ApJ...709L.146D}), and it is detected with LAT up to $\sim$$100$\,s after the
trigger.

We next estimate the total flux over a greater energy range and time interval, assuming that the observed emission around $\sim$$100$\,MeV and around $\sim$$100$\,s scale to higher energies (up to the threshold $E_{\rm CTA}$) and longer duration (out to $\sim$$1000$\,s). For the energy spectrum, we extrapolate observations at $100$\,MeV of GRB\,090510 to higher energies using the synchrotron emission by a shocked electron population with power law index $p=2.5$ (e.g., \citealt{2010ApJ...709L.146D}). For the time dependence, we extrapolate the flux of GRB\,090510 at 100\,s. We obtain the following initial flux: 
\begin{equation}
\phi^{(090510)} \approx 5\times 10^{-12}\mbox{Jy}\left(\frac{t_{\rm start}}{100\mbox{\,s}}\right)^{-1.38}\left(\frac{E_{\rm CTA}}{50\,\mbox{GeV}}\right)^{-1.25}.
\end{equation}
Starting the observations at $t_{\rm start}=100$\,s (given the time delay due to the LIGO analysis and the slewing time of CTA), and observing for $t_{\rm duration}=1000$\,s (characteristic duration of GeV-photon emission in GRB afterglows), yields a
fluence of: $\Fl\approx 4.6\times 10^{-8}\,\mbox{erg}\, \mbox{cm}^{-2}$ for $E_{\rm cutoff}=50$\,GeV and energy interval $[10^{-0.2}E_{\rm cutoff},E_{\rm cutoff}]$.

For this burst, extrapolating the $\lesssim
1$\,GeV afterglow spectrum to $50$\,GeV and beyond could be inaccurate, as for
example in a simple synchrotron interpretation the maximum synchrotron
energy is a decreasing function of time, and there may be a cutoff at these
energies. If the synchrotron emission interpretation of the afterglow is correct, we expect a cutoff at $E_{\rm cutoff}\approx 60 (\Gamma/1000)$\,GeV\footnote{This limit is obtained by equating the acceleration timescale of the radiating electrons with their cooling timescales \citep{dejager+96maxsyn}. The limit is dependent on acceleration efficiency.}. Nevertheless, the presence and value of a cutoff at this point is uncertain, and the observed data do not necessitate a cutoff.  We will therefore also consider the speculative cases when the energy spectrum continues out to $0.1$\,TeV and $1$\,TeV. We summarize the results for the cases discussed here in Table \ref{table:fluence}.

For the purposes of this study, we consider a GRB distance of 300\,Mpc
(characteristic distance at which aLIGO can detect a compact binary merger). GRB\,090510 is a uniquely bright burst, thus we
also consider GRBs with lower kinetic energy.  For fixed cutoff energy $E_{\rm cutoff}$, the flux scales with distance and energy as \citep{Granot+02Dabreaks}
\begin{equation}
\phi^{\rm syn}\propto E_{\rm kin}^{(2+p)/4} D_{\rm L}^{-2} (1+z)^{-(2+p)/4}.
\end{equation}
We only list parameters important for our study and we do not consider varying microphysical parameters.

The fluence of a 090510-like GRB at $D_{\rm L}=300$\,Mpc, observed with CTA from $t_{\rm start}=100$\,s after
trigger for a duration of $t_{\rm duration}=1000$\,s is $\Fl\approx\{3.3,2.8,1.6\}\times 10^{-5}$\,erg\,cm$^{-2}$ for the three cutoff energies. This is overwhelmingly bright, and can easily be detected by CTA.
A more likely scenario for a GRB occurring at 300\,Mpc is a burst with typical kinetic energy $E_{\rm kin}\sim 10^{51}\,$erg. The estimated fluence for this realistic case is shown in Table \ref{table:fluence}.

We also investigate the role of the time delay between the binary merger and the start of observations with CTA. To evaluate the possibility of a more delayed follow-up, we also consider an observation scenario for this typical burst for which the start of
observation is $t_{\rm start}=1000$\,s, with duration $t_{\rm duration}=1000$\,s. We also consider the possible future scenario when GW observations are performed in quasi-real time, and GW event candidate triggers are distributed to astronomers with negligible delay. For this case, we consider a time delay of $t_{\rm start}=30$\,s for the start of CTA observations. Results are shown in Table \ref{table:fluence}.

To consider a range of possible temporal delays, in Fig. \ref{figure:thresholds} we show the detectability of GRBs ($E_{\rm kin}=10^{51}$\,erg; $D_{\rm L}=300$\,Mpc) with CTA as a function of $t_{\rm start}$, with detectability defined as the ratio of GRB fluence over the detection threshold of CTA in survey mode. The results indicate that, for the lowest cutoff $E_{\rm cutoff}=50$\,GeV, GRBs are only detectable with low delays ($t_{\rm start}<10$\,s for $\sim$$1000$\,deg$^{2}$ and $t_{\rm start}<30$\,s for $\sim$$200$\,deg$^{2}$ sky area), but for higher energy thresholds, longer delays are acceptable without affecting detectability.

\subsubsection{CTA limits from self inverse Compton scattering}

For a synchrotron-emitting source we also expect a self inverse Compton component between the synchrotron photons and their emitting electrons (resulting in synchrotron self-Compton radiation, hereafter SSC). Here we
give a simplified treatment, by approximating the spectrum as joined power-law
segments (for a detailed discussion of the SSC, see \citealt{Sari+01ic}).
SSC has approximately the same spectral shape as synchrotron emission.
The peak SSC flux is scaled by the optical depth compared to the synchrotron
flux: $\phi^{\rm ssc}\approx \sigma_T n R \phi^{\rm syn}$, where $n$ is the
circumburst density, $R$ is the radius of the shock, and $\sigma_T$ is the Thomson cross section. In an adiabatically expanding shock, $R\propto t^{1/4}$ \citep{Blandford+76bm}.  We approximate the SSC spectral shape with a broken power law and neglect higher order terms \citep{Sari+01ic}.

The Klein-Nishina effect will suppress emission at the highest energies and
it may be important for $\sim$\,TeV energies \citep{Nakar+09KN}.  There are no
confirmed examples for SSC emission in short GRBs at GeV energies, although the long
lived afterglow of GRB 130427A may be interpreted as SSC emission \citep{Liu+13ic}.
Here we will focus on numerical examples to evaluate possible TeV emission from an
SSC component.

We utilize the parameters $n$, $\epsilon_B$
(fraction of energy in magnetic fields), $\epsilon_e$ (fraction of energy
carried by electrons, $E_{\rm kin}$ (kinetic energy) and electron population power-law index $p$, to describe the synchrotron component and
to calculate the SSC flux.

For realistic parameters (e.g., $n=10^{-1}$\,cm$^{-3}$, $\epsilon_{B}=10^{-1}$, $\epsilon_{e}=10^{-0.5}$, $E_{\rm
kin}=10^{51}$\,erg, and $p=2.5$) and for a GRB at $300$\,Mpc, we calculate the SSC fluence:
$\Fl\approx \{4,~6,~3\}\times 10^{-7}$\,erg\,cm$^{-2}$. For this set of
parameters the SSC emission will be detected at all three cutoff energies.  For
GWs originating from neutron star binary mergers, which might have experienced kicks form their places of birth, the
circumstellar density might be lower. Changing $n=10^{-3} \cm^{-3}$, which is roughly the halo density, in the
previous case, we get a fluence of $\Fl\approx \{3,~4,~8\}\times 10^{-9}$\,erg\,cm$^{-2}$. This lower fluence, which may be more realistic for binary mergers that left their host galaxies due to the kicks neutron stars can receive at their birth in supernovae (e.g., \citealt{2013ApJ...776...18F}), can only be detected if it extends to $\sim$\,TeV energies (and not in survey mode).

To characterize the dependence of the flux on the uncertain microphysical parameters, we
fix  the value of $p=2.5$ and get: $\phi_\nu\propto E_{\rm
kin}^{29/16}\epsilon_{B}^{7/8} \epsilon_e^{3} n^{17/16}$. This means higher
values of the equipartition parameter of the magnetic field and the electrons,
as well as the density of the interstellar material, will facilitate a detection in the TeV
band.  These scalings are valid when the TeV range falls between the injection
and characteristic frequencies of the SSC component, which is true for
parameters not too different from those presented here.

\subsubsection{Increased TeV flux from short-GRB late rebrightening}

A non-negligible fraction of short GRBs are followed by softer, extended
emission \cite{norris06}. Considering their forward shock SSC emission
in a refreshed shock \citep{Rees+98refresh} or continuous injection
\citep{Zhang+01mag}, these bursts can possibly produce a late rebrightening in
the TeV range \citep{Veres+13tev}. This increases the probability of detection
compared to simple afterglow models. The timescale of the rebrightening is
100-1000\,s, the same order of magnitude as the response time of Cherenkov
telescopes.

\subsection{Joint Detection Rates}

To estimate the rate of events that can be jointly detected by CTA and GW observatories, we consider the rate of short GRBs to be $\sim$$10$\,Gpc$^{-3}$yr$^{-1}$ (\citealt{2014MNRAS.437..649S,2012MNRAS.425.2668C,2006A&A...453..823G}; see also \citealt{2010CQGra..27q3001A}), and assume uniform source density. For a fiducial direction-averaged distance of $300$\,Mpc within which a network of GW detectors can detect a neutron star binary merger (with rotation axis pointing towards the Earth; see \citealt{2010CQGra..27q3001A}), this corresponds to $\sim$$0.3$ events per year. Taking into account a $\sim$$11\%$ duty cycle of CTA \citep{2011ExA....32..193A}, and assuming that all short GRBs within the detection horizon of GW detectors can be detected by CTA, the rate of coincident detections is $\sim$$0.03$\,yr$^{-1}$. A further decrease of $\sim 50\%$ is expected from CTA being able to observe only for source elevations $\gtrsim 30^\circ$ above the horizon. This number, nevertheless, can further increase if sub-threshold GW events with lower SNRs are followed up, or if a sub-population of short GRBs originate from black hole-neutron star mergers, which can be detected by GW observatories from larger distances. The estimate nevertheless indicates that a joint detection may require an extended period of operation.

\section{CTA follow-up survey}
\label{section:CTAsurvey}

In the previous section we estimated the detectability of short-GRBs with CTA, using a search with known source direction. In this section we estimate the same detectability, but for the case of uncertain source direction, in which CTA needs to \emph{survey} a sky area of up to $\sim$$1000$\,deg$^{2}$. For comparison, we also estimate detectability for a more accurate, $\sim$$200$\,deg$^{2}$ sky area.

CTA could carry out a fraction of its observations as \emph{''sky surveys"} \citep{2013APh....43....3A}. Following the success of the H.E.S.S. Galactic plane survey constituting 230\,hr of observations \citep{2006ApJ...636..777A}, one of the science objectives of CTA will be to carry out a survey of the inner Galactic plane. CTA will be able to do this with ten-fold improvement in sensitivity compared to H.E.S.S.  This corresponds to a uniform sensitivity down to 3\,mCrab in about the same time as the H.E.S.S. survey.

In addition, there is the possibility of a dedicated \emph{''all-sky survey"} which will observe a quarter of the sky at a sensitivity of 20\,mCrab in 370\,hr of observations \citep{2013APh....43..317D}. This all-sky survey would open the possibility of a serendipitous detection of prompt emission from short GRBs.  It is difficult to extrapolate the high-energy spectrum of a short GRB based on its low-energy emission as the high-energy spectrum has been characterized by extra spectral components above the declining low-energy flux.  If we assume that 45 short GRBs detected by Fermi-GBM ($\sim 50\%$ sky coverage) per year could all potentially have high-energy emission within the CTA sensitivity, then in survey mode there would be 9 per year all-sky assuming a $11\%$ duty cycle for CTA, or 3 during the hours devoted to this all-sky survey.  The probability of the survey patch coinciding with the GRB position is small but a serendipitous discovery is possible.

A recent review of the scientific motivation and impact of surveys with CTA \citep{2013APh....43..317D} points out that surveys have the advantage of not only generating legacy data sets, but having the potential for serendipitous discovery of TeV sources. Complementary to CTA, the High Altitude Water Cherenkov (HAWC) detector will be operating in continuous survey mode, and offers the advantage of a high duty cycle compared to imaging
atmospheric Cherenkov telescopes (IACTs), which are constrained to observe only at night time, and likely only at low Lunar illumination. HAWC has higher threshold of operation ($> 1$\,TeV) and a sensitivity goal of 1\,Crab above 1\,TeV in a day or $\sim$$50$\,mCrab in a year of operations \citep{2012NIMPA.692...72D}. While water Cherenkov detectors are excellent for transient sources and serendipitous searches, they cannot be reoriented (i.e., they cover only a fraction of the sky, $\sim$$16\%$ in the case of HAWC), and, compared to IACTs, have limited angular resolution and point source sensitivity, and operate at higher energy thresholds. CTA promises to offer a more competitive survey depth and angular resolution, with lower energy threshold for a moderate investment in observing time.

\cite{Inoue+13GRBCTA} describe planned wide-field survey CTA observations, and their potential to discover GRBs. These are not for follow up of LIGO searches, but rather for a stand-alone GRB search by divergent pointing of the telescopes of CTA to achieve a wide field of view. Such extragalactic surveys have not been carried out by IACTs before.

\begin{figure*}
\begin{center}
  \resizebox{0.8\textwidth}{!}{\includegraphics{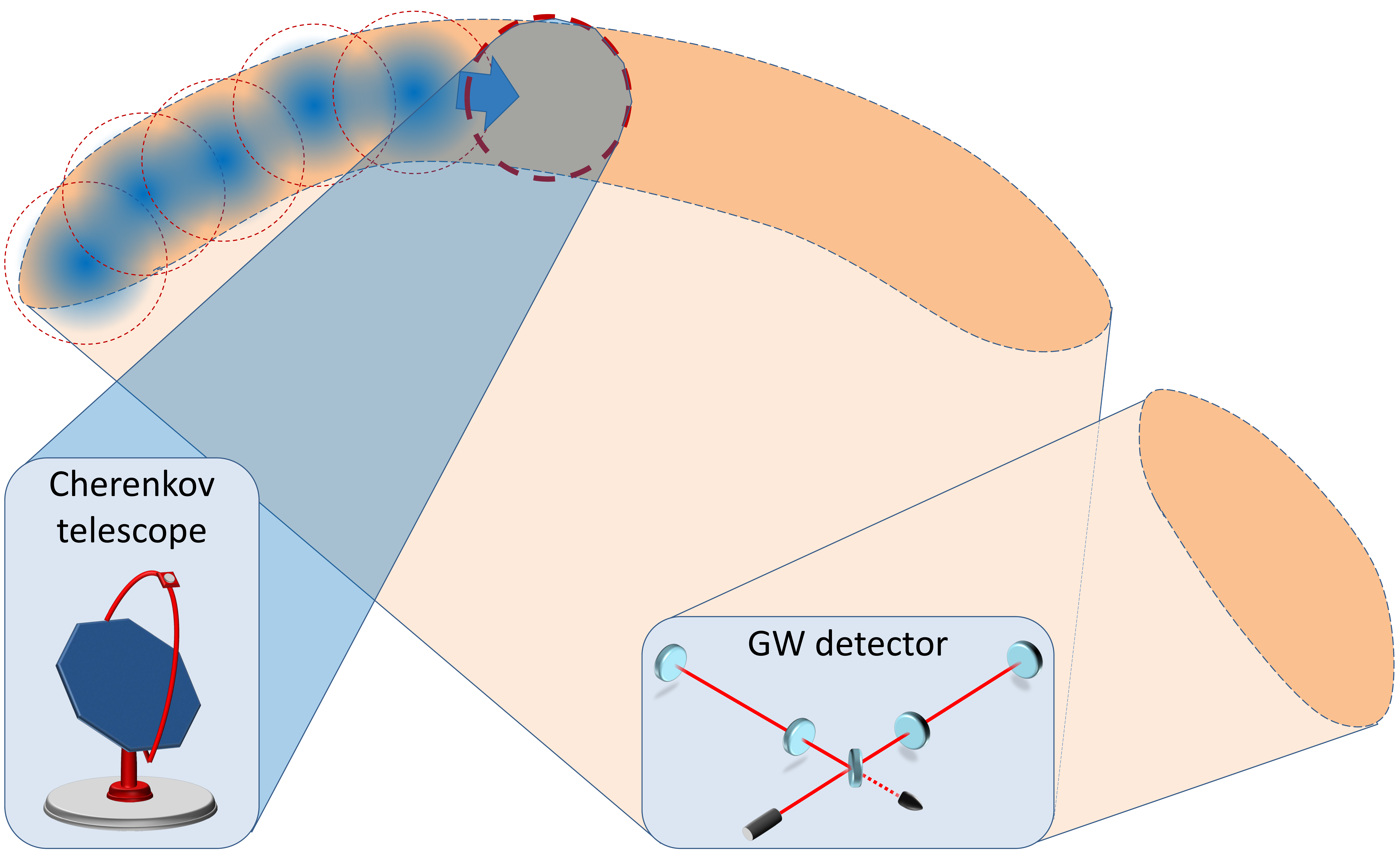}}
  \caption{Schematic representation of the sky areas of a GW event candidate and the consecutive set of sky areas covered by a CTA follow-up observation in survey mode (convergent pointing). The illustration shows a discontinuous GW sky region. Note that searches will involve multiple cherenkov telescopes and GW detectors.}\label{figure:illustration}
\end{center}
\end{figure*}

The development of the tools that allow CTA to carry out a wide-field-of-view survey will also allow for the wide-field-of-view follow-up observations presented here. A schematic drawing of the sky coverage of such a follow-up observation with CTA is shown in Fig. \ref{figure:illustration}.

\subsection{Sensitivity of CTA in survey mode}
\label{section:survey}

Here we estimate the sensitivity of CTA in a follow-up survey mode by comparing it to single-pointing observation, which is discussed above in Section \ref{section:CTAsensitivity} (hereafter single-pointing sensitivity).

A possible follow-up survey strategy with CTA is to cover the required sky area by pointing the whole telescope array (convergent pointing\footnote{Note that one may not need to use all telescopes in this mode. For example it may be sufficient to use the LSTs for this search.}) in a consecutive set of directions (see, e.g., \citealt{2013APh....43..317D}). This strategy can minimize the required software development for CTA as it relies on single-pointing observations. The sensitivity of this strategy can be directly compared to the results in Section \ref{section:CTAsensitivity}. At any given time, CTA is pointing at a given direction, therefore its sensitivity over a short time interval is the same as the single-pointing sensitivity.

The difference in the two sensitivities comes from a set of factors:

\noindent
{\bf(i)} in survey mode, CTA will point at a specific direction only for a shorter time period in order to cover the full sky area during the expected duration of multi-GeV emission. To first order, this decreases search sensitivity by a factor dependent on the fraction of the total observation time spent on each direction. The sensitivity of searching the full error region therefore changes by a factor $f_{\Omega}\simeq[(\theta_{\rm CTA}/2)^2\pi/\Omega_{\rm GW}]^{1/2}$ compared to a single-pointing survey, that is, the sky area visible to CTA at a given time ($\theta_{\rm CTA}$ is the diameter of the field of view of CTA) over the area $\Omega_{\rm GW}$ of the location error region of a GW trigger. Here we assumed that the sensitivity is background dominated, i.e., sensitivity scales with the square root of the observation duration. Further, this estimate also assumes that each direction is observed for the same duration. This will not be the case, since the expected flux from a GRB will decay with time ($\propto t^{-1.4}$, where the time $t$ since the onset of the GRB is known from the time of the GW signal). The survey will therefore need to spend time observing a given direction on the GW sky area that is proportional to $t^{3}$. 
 Here, for simplicity, we conservatively omit the effect of temporally non-uniform GRB emission.

\noindent
{\bf(ii)} Covering the GW error region with CTA tiles pointing in different directions will likely be sub-optimal. A set of CTA tiles is unlikely to exactly cover the GW error region without any overlap or overflow. A fraction of the surveyed sky area will be in directions which are not part of the GW sky area. Sub-optimal tiling will therefore introduce a factor $f_{\rm tiling}<1$ of decrease in the sensitivity of survey mode.

\noindent
{\bf(iii)} CTA has a finite slewing speed. Surveying an area larger than the field of view of CTA leads the detector to not observe in a fraction of the observation time that it spends slewing between different surveyed directions. For a total observation time $T_{\rm obs}$ and a slewing time $t_{\rm slew}$, the decrease in detection sensitivity will be $f_{\rm slew}=(1-t_{\rm slew}/T_{\rm obs})^{1/2}$, where we again assume that the sensitivity is background dominated. Taking these modifications into account, the detectable fluence threshold $S_{\rm survey}$ for the survey-mode of CTA will be
\begin{equation}
S_{\rm survey} = S_{\rm det}\,f_{\Omega}\,f_{\rm tiling}\,f_{\rm slew},
\end{equation}
where $S_{\rm det}$ is the single-pointing detection threshold of CTA.

To characterize the sensitivity of the survey mode of CTA, we estimate $S_{\rm survey}/S_{\rm det}$ using typical values for the parameters described above. Taking the field of view of CTA to be $\theta_{\rm CTA}\approx 4.6^{\circ}$ (the field of view of the large size telescopes of CTA, which are the most important at the relevant energies; \citealt{2013APh....43..317D}), and a GW error region with a total area of $1000$\,deg$^2$, we get $f_{\Omega}(1000\,\mbox{deg}^2)\approx0.13$. Similarly, a $200$\,deg$^2$ sky area gives $f_{\Omega}(200\,\mbox{deg}^2)\approx0.29$  The efficiency of tiling will depend on the shape of the GW sky area. Nevertheless, the GW sky area is unlikely to be fragmented to parts much smaller than the the field of view of CTA. Excess surveyed area will therefore come mostly from the ''edge" of the GW sky area, making this effect less significant than the decrease due to $f_{\Omega}$. Further, the sensitivity of CTA within the field of view is non-uniform, which may require partially overlapping tiling. Below we adopt $f_{\rm tiling}\approx0.75$ to account for some sensitivity decrease due to tiling. To estimate $f_{\rm slew}$, we consider a total observation time $T_{\rm obs} = 1000$\,s. For equilateral tiling (see \citealt{2013APh....43..317D}), if the GW sky area is much larger than the field of view of CTA, the characteristic total slewing angle is $\Omega_{\rm GW}/(2\cos(60^\circ)\theta_{\rm CTA}) \approx 220^\circ$, which corresponds to a slewing time of $t_{\rm slew}\lesssim25$\,s, given that the most most important large size telescopes have a slewing speed of $\sim$$20$\,s$/180^{\circ}$ \citep{2013APh....43..317D}. The slewing time is therefore negligible compared to the total observation time, even for somewhat fragmented GW sky areas. Below we consider $f_{\rm slew} \approx (T_{\rm obs} - t_{\rm slew}) / T_{\rm obs} = 0.975$. Combining these results, we arrive at
\begin{equation}
S_{\rm survey} \approx 0.1\, S_{\rm det}.
\end{equation}
We use this conversion, together with $S_{\rm det}$ obtained in Section \ref{section:CTAsensitivity}, to calculate $S_{\rm survey}$. Results are shown in Table \ref{table:fluence}.

The above estimate for the sensitivity of CTA in survey mode considers the case of background-dominated detection, i.e., in which sensitivity is determined by the signal-to-noise ratio. This will be the case for signal strengths for which the expected number of detected photons for a given pointing is $\gg 1$. Since survey-mode observations divide the full measurement time to many shorter measurements, each of these shorter measurements also have to satisfy the same criterion in order to be considered background dominated. To confirm that this will be the case, we estimated the number of detected photons for our different signal models (see Table \ref{table:fluence}), using the effective area of CTA from \cite{2013APh....43..171B}. We find that, for $n_{\rm survey}=\mathcal{O}(10)$ pointings during a survey, the number $N_\gamma^{\rm CTA}$ of photons detected from any of our GRB models with any of the considered cutoff energy thresholds will have $N_\gamma^{\rm CTA}\gg n_{\rm survey}$ for all cases in which the GRB fluence is above the detectability threshold. We therefore conclude that all cases can be considered to be background dominated.

In short, we find that the sensitivity of survey-mode searches, considering a sky area of $\sim$$1000$\,deg$^2$, is $\sim$$10\%$ of the sensitivity of single-pointing searches, while a more focused survey over $\sim$$200$\,deg$^2$ gives $\sim$$21\%$. Table \ref{table:fluence} shows that this sensitivity can still be sufficient to detect GRBs with parameters ($E_{\rm kin}=10^{51}$\,erg; $D_{\rm L}=300$\,Mpc) for emission reaching $E_{\rm cutoff}\gtrsim 100$\,GeV.

We note here that, alternatively to the convergent pointing discussed in this paper, surveys in so-called divergent mode are also possible, in which different telescopes point in different directions, therefore covering a larger part of the sky ($\sim 20^\circ \times 20^\circ$) at any given time (e.g., \citealt{2013APh....43..317D}). The possible advantages of following up GW event candidates with such divergent-mode searches using MST will be examined in a future work.




\section{Gamma-ray Burst Observations at keV-MeV photon energies}
\label{section:GBM}

Gamma-ray bursts are typically the easiest to detect in the MeV energy range where they are expected to emit the bulk of their energy output. Current instruments focusing on GRB detection in the MeV range typically have large fields of view and can efficiently detect GRBs well beyond the reach of GW detectors \citep{2005SSRv..120..143B,2009ApJ...702..791M,2011AIPC.1358..385H}. Below we examine the observations of GRBs in the MeV energy range in the context of CTA follow-up observations presented above.

Observations of the prompt MeV emission can be interesting for the purposes of GeV follow-up observations for two reasons, which we discuss below.

\subsection{Source Localization with Gamma-ray Detection}

The localization uncertainty of GRB observations is typically much smaller than the uncertainty of GW observations. If the MeV counterpart of a GW candidate is quickly identified by GRB detectors, the reconstructed location can help focus CTA observations to a smaller sky area, therefore increasing search sensitivity and allow for longer and more informative observation.

The Fermi Gamma-ray Burst Monitor (GBM; \citealt{2009ApJ...702..791M}) and the Swift Burst Alert Telescope on Swift (BAT; \citealt{2005SSRv..120..143B}) are capable of identifying GRBs within a wide field of view, and alerting other observatories with little delay.

The short-burst population detected by Swift-BAT (9 per year) may be contaminated by weak collapsar events \citep{2013ApJ...764..179B} so that the number of merger events may be smaller than 9 per year.  The overlap with CTA is thus small, but any candidate can be efficiently observed without tiling.

Fermi GBM has a duty cycle of $50\%$ for any point on the sky (Earth occultation and passage through the South Atlantic Anomaly account for the losses). Swift-BAT has a field of view of 1.4\,sr, and can localize events to within $\sim2$\,arcmin and alert external observatories with a delay $<20$\,s.

For Fermi-GBM, GRBs are localized in real-time on-board and automatically on the ground with only a few seconds latency.  The automated ground locations are within about 7.5$^\circ$ of the true position $68\%$ of the time (17$^\circ$ for $95\%$).  A refined position available within 20\,min--1\,hr after the GRB trigger is more accurate, with $68\%$ within 5$^\circ$ and $95\%$ within 10$^\circ$.  Efforts are underway to improve the real-time automated position to be of the quality of the refined position.  With 45 short GRB detections per year, Fermi-GBM could provide 2 or 3 per year above the horizon for CTA to observe with a survey tiling strategy that would be more efficient than that described in Section \ref{section:survey} \citep{Valerie}.

Additionally, the Fermi-GBM team has recently implemented an offline search for short GRBs using a new event data type that will double or triple the number of short GRBs per year.  The expected localization uncertainty will be at least 10$^\circ$ and probably larger as these are weaker events.  Because of the unknown redshift distribution of these short bursts, and of short bursts generally, it is difficult to estimate how many of these will fall within the aLIGO horizon distance.

Event candidates detected by either Swift-BAT or Fermi-GBM can be used to aid CTA follow-up observations of GW candidates either by initiating CTA follow-up observations when a short GRB triggers the instrument (if the GW candidate is not identified quickly) or by reducing the amount of sky that needs to be covered using the GW localization info alone. Nevertheless, a significant fraction ($\sim 40\%$) of nearby GRBs is not observed by either of these detectors. For these GRBs, CTA will need to rely solely on GW direction reconstruction.

The synergy with MeV GRB instruments could be valuable for GW follow-ups with CTA.  It is unclear, however, whether Swift and Fermi will be operational in the CTA era or whether any other instrument with GRB detection and real-time localization capabilities will be operation. Possible missions include UFFO \citep{2011ICRC....8..240C}, SVOM \citep{2009AIPC.1133...25G}, MIRAX \citep{2004AdSpR..34.2657B}, as well as other future ESA/NASA small missions.

\subsection{Additional Information on Source}


In some cases the MeV counterpart of a GW event candidate is identified only after the time window in which CTA follow-up observations are feasible. While, in these cases, prompt MeV observations will not help with source localization, the joint detection of GWs, prompt MeV emission and GeV emission can help us further understand the connection between the progenitor and gamma-ray emission in a wide energy band.

The interplanetary network (IPN: \cite{2011AIPC.1358..385H}), an all-sky, full-time monitor of gamma-ray transients,  is well suited for this purpose.  In its current, 9-spacecraft configuration, it detects about 325 GRBs/year, of which 18/year are short bursts.  Of the 19 short bursts with spectroscopic redshifts, the IPN has observed all events up to $z=0.45$, and $40\%$ of those with redshifts between 0.45 and 2.6.  Thus all known short GRBs at distances up to 2\,Gpc have been detected by the IPN; as the luminosity function of short bursts is not known, however, it is conceivable that some weak events could go undetected.  Their number cannot be estimated.  The sensitivities and energy ranges of the individual detectors aboard the spacecraft vary considerably from one experiment to the next, but the overall IPN sensitivity can be characterized by a fluence of $\sim10^{-6}$\,erg\,cm$^{-2}$, and/or a peak flux of $\sim1$photon\,cm$^{-2}$\,s$^{-1}$, in the 25-150\,keV energy range.  GRBs above these levels have a $50\%$ chance or greater of being detected by at least two spacecraft in the network.

The network presently consists of 5 spacecraft in near-Earth orbit, two at distances up to about 5 light-seconds from Earth, and two in orbit around Mercury and Mars.  Thus, when the duty cycles and planet-blocking constraints of the network as a whole are considered, the entire sky is viewed without interruption.  The IPN localizes bursts by triangulation (i.e. arrival time analysis), and the error box dimensions have a broad distribution from arcminutes to 10’s of degrees, depending upon the GRB intensity and the number of spacecraft which observed it.  The delays to obtain localizations range from hours in the best cases to a few days in the worst cases.  Given the detection rates and localization areas, a temporal and spatial coincidence between an IPN GRB and a GW observation would be highly significant in almost all cases, and would considerably strengthen both the case for the reality of a GW detection, and its identification as a cosmic gamma-ray burst.  Indeed, the IPN and LIGO teams have worked together since LIGO's earliest engineering runs.

The configuration of the IPN changes continually as old missions are retired and new ones replace them.  While it is virtually certain that some near-Earth missions will be retired in the near future, the exact configuration in the advanced LIGO era is unpredictable.  Some near-Earth missions will be replaced by new missions; others will not.  The reduction of the number of near-Earth spacecraft to 2 or 3, however, would have relatively little impact, given their redundancy.  The fates of the missions farther from Earth depend on funding, as well as on their utility for the scientific objectives for which they were designed (only one is an astrophysics mission).  It is conceivable that both planetary missions will be taken out of service, but that at least one new mission will come on-line.  This would have the effect of truncating the distribution of error box areas below a few degrees.  Nevertheless, the probability of a random spatial/temporal coincidence between a GW event and a GRB would still be sufficiently small to be very significant.

\section{Conclusion}
\label{section:conclusion}

We explored the feasibility of following up GW events with CTA. We focused on the scenario in which the GW event is poorly localized, necessitating follow-up observations to cover up to $\sim$$1000$\,deg$^2$ of sky area. Limited localization can emerge from various detection scenarios. In the early advanced GW detector era, one can expect only the two LIGO observatories to operate at high sensitivity, and direction reconstruction with two detectors is limited. But even with further GW detectors in operation, GW event candidates with relatively low signal-to-noise ratios will also have poorly constrained directions of origin, therefore requiring follow-up over a larger sky area.

We based our study on short GRBs, assuming that they originate from compact binary mergers, which are considered the most promising sources for the first GW detections.

While various follow-up observations (e.g., optical/infrared) will be difficult to carry out over larger ($\gg 100$\,deg$^2$) sky areas with the desired sensitivity, we find that CTA may be capable of efficiently detecting late-time high-energy gamma-ray emission from short GRBs. To estimate their detectability, we extrapolated the energy spectrum observed by Fermi-LAT to $\gtrsim 50$\,GeV where CTA becomes sensitive. Currently it is unclear, due to the lack of observations, whether short-GRB spectra extend into this range, and to how high an energy. We considered different cutoff energies (from 50\,GeV to 1\,TeV), as well as multiple GRB emission scenarios, to investigate the sensitivity of CTA for these different cases.

Our results show that short GRBs with high-energy emission extending up to $\sim100$\,GeV can be detectable via CTA, even if CTA needs to survey a sky area of $\sim$$1000$\,deg$^2$ and if CTA observations are delayed by $\sim 100$\,s following the onset of gamma-ray emission. Detection with lower energy cutoffs is also promising, although may require a dense circumburst medium, faster GW event reconstruction, smaller sky area, or closer source. For comparison we considered a $\sim$$200$\,deg$^2$ sky area that can be achieved for some events with stronger GW emission, or if we restrict our search to a fraction of the sky area with the highest-probability directions. For a $\sim$$200$\,deg$^2$ sky area, we find that GRBs even with cutoffs somewhat below $100$\,GeV can be detectable, although for a $\sim50$\,GeV cutoff one requires faster response than $\sim 100$\,s.

Many of the events detected by both GW facilities and CTA will also likely be observed by GRB satellites. For observations with low latency, as in the case of Fermi-GBM and Swift-BAT, the localization of the GRB can significantly reduce the sky area CTA needs to cover in order to find the source. The detection of MeV emission can also be important in mapping the connection between GW and electromagnetic emission within a broad energy range.

We estimated the rate of events that can be jointly detected by CTA and GW observatories, consider a characteristic short-GRB rate of $10$\,Gpc$^{-3}$yr$^{-1}$, and a fiducial GW horizon distance of of $300$\,Mpc. With $\sim$$11\%$ duty cycle for CTA we find a limited detection rate of $\sim$$0.03$\,yr$^{-1}$. A further decrease of $\sim 50\%$ is expected from CTA being able to observe only for source elevations $\gtrsim 30^\circ$ above the horizon. This number, nevertheless, can increase if sub-threshold GW events with lower SNRs are followed up, or if a sub-population of short GRBs originate from black hole-neutron star mergers, which can be detected by GW observatories from larger distances. The estimate nevertheless indicates that a joint detection may require an extended period of operation.

Overall, we find that CTA is well suited to perform follow-up observations of GW events, even those with limited source localization. It can, therefore, be important to carry out a more detailed investigation of the possible follow-up observation strategies with CTA, and the expected joint sensitivity, beginning as early as in the installation phase. It will also be important to further our understanding of the phenomenology of GRB emission at $\gg$\,GeV energies.

\section*{Acknowledgments}

The authors thank Neil Gehrels, Brian Metzger and Leo Singer for useful comments. This work was approved for publication by the LIGO Scientific Collaboration. JPO acknowledges the support of the UK Space Agency. IB and SM are thankful for the generous support of Columbia University in the City of New York and the National Science Foundation under cooperative agreement PHY-0847182.

\bsp

\label{lastpage}


\begin{thebibliography}{1}
\expandafter\ifx\csname natexlab\endcsname\relax\def\natexlab#1{#1}\fi

\bibitem[{{Abadie} {et~al.}(2010){Abadie}, {Abbott}, {Abbott}, {Abernathy},
  {Accadia}, {Acernese}, {Adams}, {Adhikari}, {Ajith}, {Allen}, \&
  et~al.}]{2010CQGra..27q3001A}
{Abadie}, J., {Abbott}, B.~P., {Abbott}, R., {et~al.} 2010, CQG, 27, 173001

\bibitem[{{Abadie} {et~al.}(2012{\natexlab{a}}){Abadie}, {Abbott}, {Abbott},
  {Abbott}, {Abernathy}, {Accadia}, {Acernese}, {Adams}, {Adhikari}, {Affeldt},
  \& et~al.}]{2012A&A...541A.155A}
---. 2012{\natexlab{a}}, \aap, 541, A155

\bibitem[{{Abadie} {et~al.}(2012{\natexlab{b}}){Abadie}, {Abbott}, {Abbott},
  {Abbott}, {Abernathy}, {Accadia}, {Acernese}, {Adams}, {Adhikari}, {Affeldt},
  \& et~al.}]{2012PhRvD..85h2002A}
---. 2012{\natexlab{b}}, \prd, 85, 082002

\bibitem[{{Abdo} \& the {Fermi}~collaboration(2009)}]{Abdo+09-090902B}
{Abdo}, A.~A., \& the {Fermi}~collaboration. 2009, \apjl, 706, L138

\bibitem[{{Abdo} {et~al.}(2009){Abdo}, {Ackermann}, {Ajello}, {Asano},
  {Atwood}, {Axelsson}, {Baldini}, {Ballet}, {Barbiellini}, {Baring},
  {Bastieri}, {Bechtol}, {Bellazzini}, {Berenji}, {Bhat}, {Bissaldi}, {Bloom},
  {Bonamente}, {Bonnell}, {Borgland}, {Bouvier}, {Bregeon}, {Brez}, {Briggs},
  {Brigida}, {Bruel}, {Burgess}, {Burnett}, {Caliandro}, {Cameron}, {Caraveo},
  {Casandjian}, {Cecchi}, {{\c C}elik}, {Chaplin}, {Charles}, {Cheung},
  {Chiang}, {Ciprini}, {Claus}, {Cohen-Tanugi}, {Cominsky}, {Connaughton},
  {Conrad}, {Cutini}, {Dermer}, {de Angelis}, {de Palma}, {Digel}, {Dingus},
  {Do Couto E Silva}, {Drell}, {Dubois}, {Dumora}, {Farnier}, {Favuzzi},
  {Fegan}, {Finke}, {Fishman}, {Focke}, {Foschini}, {Fukazawa}, {Funk},
  {Fusco}, {Gargano}, {Gasparrini}, {Gehrels}, {Germani}, {Gibby}, {Giebels},
  {Giglietto}, {Giordano}, {Glanzman}, {Godfrey}, {Granot}, {Greiner},
  {Grenier}, {Grondin}, {Grove}, {Grupe}, {Guillemot}, {Guiriec}, {Hanabata},
  {Harding}, {Hayashida}, {Hays}, {Hoversten}, {Hughes}, {J{\'o}hannesson},
  {Johnson}, {Johnson}, {Johnson}, {Kamae}, {Katagiri}, {Kataoka}, {Kawai},
  {Kerr}, {Kippen}, {Kn{\"o}dlseder}, {Kocevski}, {Kouveliotou}, {Kuehn},
  {Kuss}, {Lande}, {Latronico}, {Lemoine-Goumard}, {Longo}, {Loparco}, {Lott},
  {Lovellette}, {Lubrano}, {Madejski}, {Makeev}, {Mazziotta}, {McBreen},
  {McEnery}, {McGlynn}, {M{\'e}sz{\'a}ros}, {Meurer}, {Michelson},
  {Mitthumsiri}, {Mizuno}, {Moiseev}, {Monte}, {Monzani}, {Moretti},
  {Morselli}, {Moskalenko}, {Murgia}, {Nakamori}, {Nolan}, {Norris}, {Nuss},
  {Ohno}, {Ohsugi}, {Omodei}, {Orlando}, {Ormes}, {Ozaki}, {Paciesas},
  {Paneque}, {Panetta}, {Parent}, {Pelassa}, {Pepe}, {Pesce-Rollins},
  {Petrosian}, {Piron}, {Porter}, {Preece}, {Rain{\`o}}, {Ramirez-Ruiz},
  {Rando}, {Razzano}, {Razzaque}, {Reimer}, {Reimer}, {Reposeur}, {Ritz},
  {Rochester}, {Rodriguez}, {Roth}, {Ryde}, {Sadrozinski}, {Sanchez}, {Sander},
  {Saz Parkinson}, {Scargle}, {Schalk}, {Sgr{\`o}}, {Siskind}, {Smith},
  {Smith}, {Spandre}, {Spinelli}, {Stamatikos}, {Stecker}, {Strickman},
  {Suson}, {Tajima}, {Takahashi}, {Takahashi}, {Tanaka}, {Thayer}, {Thayer},
  {Thompson}, {Tibaldo}, {Toma}, {Torres}, {Tosti}, {Troja}, {Uchiyama},
  {Uehara}, {Usher}, {van der Horst}, {Vasileiou}, {Vilchez}, {Vitale}, {von
  Kienlin}, {Waite}, {Wang}, {Wilson-Hodge}, {Winer}, {Wood}, {Wu}, {Yamazaki},
  {Ylinen}, {Ziegler}, \& {the Fermi LAT Collaboration}}]{Abdo+09-090510}
{Abdo}, A.~A., {Ackermann}, M., {Ajello}, M., {et~al.} 2009, \nat, 462, 331

\bibitem[{{Abdo} {et~al.}(2010{\natexlab{a}}){Abdo}, {Ackermann}, {Ajello},
  {Asano}, {Atwood}, {Axelsson}, {Baldini}, {Ballet}, {Barbiellini},
  {Bastieri}, {Baughman}, {Bechtol}, {Bellazzini}, {Berenji}, {Bhat},
  {Bissaldi}, {Blandford}, {Bloom}, {Bonamente}, {Borgland}, {Bouvier},
  {Bregeon}, {Brez}, {Briggs}, {Brigida}, {Bruel}, {Burgess}, {Burnett},
  {Buson}, {Caliandro}, {Cameron}, {Caraveo}, {Carrigan}, {Casandjian},
  {Cecchi}, {{\c C}elik}, {Chaplin}, {Charles}, {Chekhtman}, {Chiang},
  {Ciprini}, {Claus}, {Cohen-Tanugi}, {Cominsky}, {Connaughton}, {Conrad},
  {Cutini}, {Dermer}, {de Angelis}, {de Palma}, {Digel}, {Silva}, {Drell},
  {Dubois}, {Dumora}, {Farnier}, {Favuzzi}, {Fegan}, {Fishman}, {Focke},
  {Fortin}, {Frailis}, {Fukazawa}, {Funk}, {Fusco}, {Gargano}, {Gasparrini},
  {Gehrels}, {Germani}, {Giebels}, {Giglietto}, {Giommi}, {Giordano},
  {Glanzman}, {Godfrey}, {Granot}, {Grenier}, {Grondin}, {Grove}, {Guillemot},
  {Guiriec}, {Hanabata}, {Harding}, {Hayashida}, {Haynes}, {Hays}, {Horan},
  {Hughes}, {Jackson}, {J{\'o}hannesson}, {Johnson}, {Johnson}, {Kamae},
  {Katagiri}, {Kataoka}, {Kawai}, {Kerr}, {Kippen}, {Kn{\"o}dlseder},
  {Kocevski}, {Kocian}, {Komin}, {Kouveliotou}, {Kuehn}, {Kuss}, {Lande},
  {Latronico}, {Lemoine-Goumard}, {Longo}, {Loparco}, {Lott}, {Lovellette},
  {Lubrano}, {Madejski}, {Makeev}, {Mazziotta}, {McBreen}, {McEnery},
  {McGlynn}, {Meegan}, {M{\'e}sz{\'a}ros}, {Meurer}, {Michelson},
  {Mitthumsiri}, {Mizuno}, {Moiseev}, {Monte}, {Monzani}, {Moretti},
  {Morselli}, {Moskalenko}, {Murgia}, {Nakamori}, {Nolan}, {Norris}, {Nuss},
  {Ohno}, {Ohsugi}, {Omodei}, {Orlando}, {Ormes}, {Paciesas}, {Paneque},
  {Panetta}, {Parent}, {Pelassa}, {Pepe}, {Pesce-Rollins}, {Piron}, {Porter},
  {Preece}, {Rain{\`o}}, {Rando}, {Razzano}, {Razzaque}, {Reimer}, {Reimer},
  {Reposeur}, {Ripken}, {Ritz}, {Rochester}, {Rodriguez}, {Roth}, {Ryde},
  {Sadrozinski}, {Sanchez}, {Sander}, {Saz Parkinson}, {Scargle}, {Schalk},
  {Sgr{\`o}}, {Siskind}, {Smith}, {Smith}, {Spandre}, {Spinelli}, {Stamatikos},
  {Strickman}, {Suson}, {Tagliaferri}, {Tajima}, {Takahashi}, {Tanaka},
  {Thayer}, {Thayer}, {Thompson}, {Tibaldo}, {Toma}, {Torres}, {Tosti},
  {Tramacere}, {Troja}, {Uchiyama}, {Usher}, {van der Horst}, {Vasileiou},
  {Vilchez}, {Vitale}, {von Kienlin}, {Waite}, {Wang}, {Wilson-Hodge}, {Winer},
  {Wood}, {Wu}, {Yamazaki}, {Ylinen}, \& {Ziegler}}]{2010ApJ...712..558A}
---. 2010{\natexlab{a}}, \apj, 712, 558

\bibitem[{{Abdo} {et~al.}(2010{\natexlab{b}}){Abdo}, {Ackermann}, {Ajello},
  {Allafort}, {Antolini}, {Atwood}, {Axelsson}, {Baldini}, {Ballet},
  {Barbiellini}, \& et~al.}]{2010ApJS..188..405A}
---. 2010{\natexlab{b}}, \apjs, 188, 405

\bibitem[{{Abdo} {et~al.}(2012){Abdo}, {Abeysekara}, {Allen}, {Aune}, {Berley},
  {Chen}, {Christopher}, {DeYoung}, {Dingus}, {Ellsworth}, {Gonzalez},
  {Goodman}, {Granot}, {Hays}, {Hoffman}, {H{\"u}ntemeyer}, {Kolterman},
  {Linnemann}, {McEnery}, {Mincer}, {Morgan}, {Nemethy}, {Pretz},
  {Ramirez-Ruiz}, {Ryan}, {Saz Parkinson}, {Shoup}, {Sinnis}, {Smith},
  {Vasileiou}, {Walker}, {Williams}, \& {Yodh}}]{Abdo+12milagro}
{Abdo}, A.~A., {Abeysekara}, A.~U., {Allen}, B.~T., {et~al.} 2012, \apjl, 753,
  L31

\bibitem[{{Acciari} {et~al.}(2011){Acciari}, {Aliu}, {Arlen}, {Aune},
  {Beilicke}, {Benbow}, {Bradbury}, {Buckley}, {Bugaev}, {Byrum}, {Cannon},
  {Cesarini}, {Christiansen}, {Ciupik}, {Collins-Hughes}, {Connolly}, {Cui},
  {Duke}, {Errando}, {Falcone}, {Finley}, {Finnegan}, {Fortson}, {Furniss},
  {Galante}, {Gall}, {Godambe}, {Griffin}, {Grube}, {Guenette}, {Gyuk},
  {Hanna}, {Holder}, {Hughes}, {Hui}, {Humensky}, {Jackson}, {Kaaret},
  {Karlsson}, {Kertzman}, {Kieda}, {Krawczynski}, {Krennrich}, {Lang},
  {Madhavan}, {Maier}, {McArthur}, {McCann}, {Moriarty}, {Newbold}, {Ong},
  {Orr}, {Otte}, {Park}, {Perkins}, {Pohl}, {Prokoph}, {Quinn}, {Ragan},
  {Reyes}, {Reynolds}, {Roache}, {Rose}, {Ruppel}, {Saxon}, {Schroedter},
  {Sembroski}, {{\c S}ent{\"u}rk}, {Smith}, {Staszak}, {Swordy}, {Te{\v
  s}i{\'c}}, {Theiling}, {Thibadeau}, {Tsurusaki}, {Varlotta}, {Vassiliev},
  {Vincent}, {Vivier}, {Wakely}, {Ward}, {Weekes}, {Weinstein}, {Weisgarber},
  {Williams}, \& {Wood}}]{2011ApJ...743...62A}
{Acciari}, V.~A., {Aliu}, E., {Arlen}, T., {et~al.} 2011, \apj, 743, 62

\bibitem[{Acharya {et~al.}(2013)Acharya, Actis, Aghajani, Agnetta, Aguilar,
  Aharonian, Ajello, Akhperjanian, Alcubierre, Aleksić, Alfaro, Aliu,
  Allafort, Allan, Allekotte, Amato, Anderson, Angüner, Antonelli, Antoranz,
  Aravantinos, Arlen, Armstrong, Arnaldi, Arrabito, Asano, Ashton, Asorey,
  Awane, Baba, Babic, Baby, Bähr, Bais, Baixeras, Bajtlik, Balbo, Balis,
  Balkowski, Bamba, Bandiera, Barber, Barbier, Barceló, Barnacka, Barnstedt,
  de~Almeida, Barrio, Basili, Basso, Bastieri, Bauer, Baushev, Becerra,
  Becherini, Bechtol, Tjus, Beckmann, Bednarek, Behera, Belluso, Benbow,
  Berdugo, Berger, Bernard, Bernardino, Bernlöhr, Bhat, Bhattacharyya,
  Bigongiari, Biland, Billotta, Bird, Birsin, Bissaldi, Biteau, Bitossi, Blake,
  Bigas, Blasi, Bobkov, Boccone, Boettcher, Bogacz, Bogart, Bogdan, Boisson,
  Gargallo, Bolmont, Bonanno, Bonardi, Bonev, Bonifacio, Bonnoli, Bordas,
  Borgland, Borkowski, Bose, Botner, Bottani, Bouchet, Bourgeat, Boutonnet,
  Bouvier, Brau-Nogué, Braun, Bretz, Briggs, Bringmann, Brook, Brun, Brunetti,
  Buanes, Buckley, Buehler, Bugaev, Bulgarelli, Bulik, Busetto, Buson, Byrum,
  Cailles, Cameron, Camprecios, Canestrari, Cantu, Capalbi, Caraveo, Carmona,
  Carosi, Carr, Carton, Casanova, Casiraghi, Catalano, Cavazzani, Cazaux,
  Cerruti, Chabanne, Chadwick, Champion, Chen, Chiang, Chiappetti, Chikawa,
  Chitnis, Chollet, Chudoba, Cieślar, Cillis, Cohen-Tanugi, Colafrancesco,
  Colin, Colome, Colonges, Compin, Conconi, Conforti, Connaughton, Conrad,
  Contreras, Coppi, Corona, Corti, Cortina, Cossio, Costantini, Cotter, Courty,
  Couturier, Covino, Crimi, Criswell, Croston, Cusumano, Dafonseca, Dale,
  Daniel, Darling, Davids, Dazzi, Angelis, Caprio, Frondat, de~Gouveia
  Dal~Pino, de~la Calle, Vega, de~los Reyes~Lopez, Lotto, Luca, de~Mello~Neto,
  de~Naurois, de~Oliveira, de~Oña~Wilhelmi, de~Souza, Decerprit, Decock, Deil,
  Delagnes, Deleglise, Delgado, Volpe, Demange, Depaola, Dettlaff, Paola,
  Pierro, Díaz, Dick, Dickherber, Dickinson, Diez-Blanco, Digel, Dimitrov,
  Disset, Djannati-Ataï, Doert, Dohmke, Domainko, Prester, Donat, Dorner,
  Doro, Dournaux, Drake, Dravins, Drury, Dubois, Dubois, Dubus, Dufour, Dumas,
  Dumm, Durand, Dyks, Dyrda, Ebr, Edy, Egberts, Eger, Einecke, Eleftheriadis,
  Elles, Emmanoulopoulos, Engelhaupt, Enomoto, Ernenwein, Errando, Etchegoyen,
  Evans, Falcone, Fantinel, Farakos, Farnier, Fasola, Favill, Fede, Federici,
  Fegan, Feinstein, Ferenc, Ferrando, Fesquet, Fiasson, Fillin-Martino, Fink,
  Finley, Finley, Fiorini, Curcoll, Flores, Florin, Focke, Föhr, Fokitis,
  Font, Fontaine, Fornasa, Förster, Fortson, Fouque, Franckowiak, Fransson,
  Fraser, Frei, Albuquerque, Fresnillo, Fruck, Fujita, Fukazawa, Fukui, Funk,
  Gäbele, Gabici, Gabriele, Gadola, Galante, Gall, Gallant, Gámez-García,
  García, López, Gardiol, Garrido, Garrido, Gascon, Gaug, Gaweda,
  Gebremedhin, Geffroy, Gerard, Ghedina, Ghigo, Giannakaki, Gianotti,
  Giarrusso, Giavitto, Giebels, Gika, Giommi, Girard, Giro, Giuliani, Glanzman,
  Glicenstein, Godinovic, Golev, Berisso, Gómez-Ortega, Gonzalez, González,
  González, Muñoz, Gothe, Gougerot, Graciani, Grandi, Grañena, Granot,
  Grasseau, Gredig, Green, Greenshaw, Grégoire, Grimm, Grube, Grudzinska,
  Gruev, Grünewald, Grygorczuk, Guarino, Gunji, Gyuk, Hadasch, Hagiwara, Hahn,
  Hakansson, Hallgren, Heras, Hara, Hardcastle, Harris, Hassan, Hatanaka,
  Haubold, Haupt, Hayakawa, Hayashida, Heller, Henault, Henri, Hermann, Hermel,
  Herrero, Hidaka, Hinton, Hoffmann, Hofmann, Hofverberg, Holder, Horns,
  Horville, Houles, Hrabovsky, Hrupec, Huan, Huber, Huet, Hughes, Humensky,
  Huovelin, Ibarra, Illa, Impiombato, Incorvaia, Inoue, Inoue, Ioka, Ismailova,
  Jablonski, Jacholkowska, Jamrozy, Janiak, Jean, Jeanney, Jimenez, Jogler,
  Johnson, Journet, Juffroy, Jung, Kaaret, Kabuki, Kagaya, Kakuwa, Kalkuhl,
  Kankanyan, Karastergiou, Kärcher, Karczewski, Karkar, Kasperek, Kastana,
  Katagiri, Kataoka, Katarzyński, Katz, Kawanaka, Kellner-Leidel, Kelly,
  Kendziorra, Khélifi, Kieda, Kifune, Kihm, Kishimoto, Kitamoto, Kluźniak,
  Knapic, Knapp, Knödlseder, Köck, Kocot, Kodani, Köhne, Kohri, Kokkotas,
  Kolitzus, Komin, Kominis, Konno, Köppel, Korohoda, Kosack, Koss,
  Kossakowski, Kostka, Koul, Kowal, Koyama, Kozioł, Krähenbühl, Krause,
  Krawzcynski, Krennrich, Krepps, Kretzschmann, Krobot, Krueger, Kubo,
  Kudryavtsev, Kushida, Kuznetsov, Barbera, Palombara, Parola, Rosa, Lacombe,
  Lamanna, Lande, Languignon, Lapington, Laporte, Lavalley, Flour, Padellec,
  Lee, Lee, de~Oliveira, Lelas, Lenain, Leopold, Lerch, Lessio, Lieunard,
  Lindfors, Liolios, Lipniacka, Lockart, Lohse, Lombardi, Lopatin, Lopez,
  López-Coto, López-Oramas, Lorca, Lorenz, Lubinski, Lucarelli, Lüdecke,
  Ludwin, Luque-Escamilla, Lustermann, Luz, Lyard, Maccarone, Maccarone,
  Madejski, Madhavan, Mahabir, Maier, Majumdar, Malaguti, Maltezos, Manalaysay,
  Mancilla, Mandat, Maneva, Mangano, Manigot, Mannheim, Manthos, Maragos,
  Marcowith, Mariotti, Marisaldi, Markoff, Marszałek, Martens, Martí, Martin,
  Martin, Martínez, Martínez, Martínez, Masserot, Mastichiadis, Mathieu,
  Matsumoto, Mattana, Mattiazzo, Maurin, Maxfield, Maya, Mazin, Comb, McCubbin,
  McHardy, McKay, Medina, Melioli, Melkumyan, Mereghetti, Mertsch, Meucci,
  Michałowski, Micolon, Mihailidis, Mineo, Minuti, Mirabal, Mirabel, Miranda,
  Mirzoyan, Mizuno, Moal, Moderski, Mognet, Molinari, Molinaro, Montaruli,
  Monteiro, Moore, Olaizola, Mordalska, Morello, Mori, Mottez, Moudden, Moulin,
  Mrusek, Mukherjee, Munar-Adrover, Muraishi, Murase, Murphy, Nagataki, Naito,
  Nakajima, Nakamori, Nakayama, Naumann, Naumann, Naumann-Godo, Nayman, Nedbal,
  Neise, Nellen, Neustroev, Neyroud, Nicastro, Nicolau-Kukliński,
  Niedźwiecki, Niemiec, Nieto, Nikolaidis, Nishijima, Nolan, Northrop, Nosek,
  Nowak, Nozato, O’Brien, Ohira, Ohishi, Ohm, Ohoka, Okuda, Okumura, Olive,
  Ong, Orito, Orr, Osborne, Ostrowski, Otero, Otte, Ovcharov, Oya, Ozieblo,
  Padilla, Paiano, Paillot, Paizis, Palanque, Palatka, Pallota, Panagiotidis,
  Panazol, Paneque, Panter, Paoletti, Papayannis, Papyan, Paredes, Pareschi,
  Parks, Parraud, Parsons, Arribas, Pech, Pedaletti, Pelassa, Pelat, Perez,
  Persic, Petrucci, Peyaud, Pichel, Pita, Pizzolato, Platos, Platzer, Pogosyan,
  Pohl, Pojmanski, Ponz, Potter, Poutanen, Prandini, Prast, Preece, Profeti,
  Prokoph, Prouza, Proyetti, Puerto-Gimenez, Pühlhofer, Puljak, Punch,
  Pyzioł, Quel, Quinn, Quirrenbach, Racero, Rajda, Ramon, Rando, Rannot,
  Rataj, Raue, Reardon, Reimann, Reimer, Reimer, Reitberger, Renaud, Renner,
  Reville, Rhode, Ribó, Ribordy, Richer, Rico, Ridky, Rieger, Ringegni,
  Ripken, Ristori, Riviére, Rivoire, Rob, Roeser, Rohlfs, Rojas, Romano,
  Romaszkan, Romero, Rosen, Lees, Ross, Rouaix, Rousselle, Rousselle, Rovero,
  Roy, Royer, Rudak, Rulten, Rupiński, Russo, Ryde, Sacco, Saemann, Saggion,
  Sahakian, Saito, Saito, Saito, Sakaki, Sakonaka, Salini, Sanchez,
  Sanchez-Conde, Sandoval, Sandaker, Sant’Ambrogio, Santangelo, Santos,
  Sanuy, Sapozhnikov, Sarkar, Sartore, Sasaki, Satalecka, Sawada, Scalzotto,
  Scapin, Scarcioffolo, Schafer, Schanz, Schlenstedt, Schlickeiser, Schmidt,
  Schmoll, Schovanek, Schroedter, Schultz, Schultze, Schulz, Schure, Schwab,
  Schwanke, Schwarz, Schwarzburg, Schweizer, Schwemmer, Segreto, Seiradakis,
  Sembroski, Seweryn, Sharma, Shayduk, Shellard, Shi, Shibata, Shibuya, Shum,
  Sidoli, Sidz, Sieiro, Sikora, Silk, Sillanpää, Singh, Sitarek, Skole,
  Smareglia, Smith, Smith, Smith, Smith, Sobczyńska, Sol, Sottile, Sowiński,
  Spanier, Spiga, Spyrou, Stamatescu, Stamerra, Starling, Stawarz, Steenkamp,
  Stegmann, Steiner, Stergioulas, Sternberger, Sterzel, Stinzing, Stodulski,
  Straumann, Strazzeri, Stringhetti, Suarez, Suchenek, Sugawara, Sulanke, Sun,
  Supanitsky, Suric, Sutcliffe, Sykes, Szanecki, Szepieniec, Szostek,
  Tagliaferri, Tajima, Takahashi, Takahashi, Takalo, Takami, Talbot, Tammi,
  Tanaka, Tanaka, Tasan, Tavani, Tavernet, Tejedor, Telezhinsky, Temnikov,
  Tenzer, Terada, Terrier, Teshima, Testa, Tezier, Thuermann, Tibaldo, Tibolla,
  Tiengo, Tluczykont, Peixoto, Tokanai, Tokarz, Toma, Torii, Tornikoski,
  Torres, Torres, Tosti, Totani, Toussenel, Tovmassian, Travnicek, Trifoglio,
  Troyano, Tsinganos, Ueno, Umehara, Upadhya, Usher, Uslenghi, Valdes-Galicia,
  Vallania, Vallejo, van Driel, van Eldik, Vandenbrouke, Vanderwalt, Vankov,
  Vasileiadis, Vassiliev, Veberic, Vegas, Vercellone, Vergani, Veyssiére,
  Vialle, Viana, Videla, Vincent, Vincent, Vink, Vlahakis, Vlahos, Vogler,
  Vollhardt, von Gunten, Vorobiov, Vuerli, Waegebaert, Wagner, Wagner, Wagner,
  Wakely, Walter, Walther, Warda, Warwick, Wawer, Wawrzaszek, Webb, Wegner,
  Weinstein, Weitzel, Welsing, Werner, Wetteskind, White, Wierzcholska,
  Wiesand, Wilkinson, Williams, Willingale, Winiarski, Wischnewski,
  Wiśniewski, Wood, Wörnlein, Xiong, Yadav, Yamamoto, Yamamoto, Yamazaki,
  Yanagita, Yebras, Yelos, Yoshida, Yoshida, Yoshikoshi, Zabalza, Zacharias,
  Zajczyk, Zanin, Zdziarski, Zech, Zhao, Zhou, Ziętara, Ziolkowski,
  Ziółkowski, Zitelli, Zurbach, \& Żychowski}]{Acharya20133}
Acharya, B., Actis, M., Aghajani, T., {et~al.} 2013, Astroparticle Physics, 43,
  3 , seeing the High-Energy Universe with the Cherenkov Telescope Array - The
  Science Explored with the \{CTA\}

\bibitem[{{Acharya} {et~al.}(2013){Acharya}, {Actis}, {Aghajani}, {Agnetta},
  {Aguilar}, {Aharonian}, {Ajello}, {Akhperjanian}, {Alcubierre},
  {Aleksi{\'c}}, \& et~al.}]{2013APh....43....3A}
{Acharya}, B.~S., {Actis}, M., {Aghajani}, T., {et~al.} 2013, Astroparticle
  Physics, 43, 3

\bibitem[{{Ackermann} {et~al.}(2010){Ackermann}, {Asano}, {Atwood}, {Axelsson},
  {Baldini}, {Ballet}, {Barbiellini}, {Baring}, {Bastieri}, {Bechtol},
  {Bellazzini}, {Berenji}, {Bhat}, {Bissaldi}, {Blandford}, {Bloom},
  {Bonamente}, {Borgland}, {Bouvier}, {Bregeon}, {Brez}, {Briggs}, {Brigida},
  {Bruel}, {Buson}, {Caliandro}, {Cameron}, {Caraveo}, {Carrigan},
  {Casandjian}, {Cecchi}, {{\c C}elik}, {Charles}, {Chiang}, {Ciprini},
  {Claus}, {Cohen-Tanugi}, {Connaughton}, {Conrad}, {Dermer}, {de Palma},
  {Dingus}, {Silva}, {Drell}, {Dubois}, {Dumora}, {Farnier}, {Favuzzi},
  {Fegan}, {Finke}, {Focke}, {Frailis}, {Fukazawa}, {Fusco}, {Gargano},
  {Gasparrini}, {Gehrels}, {Germani}, {Giglietto}, {Giordano}, {Glanzman},
  {Godfrey}, {Granot}, {Grenier}, {Grondin}, {Grove}, {Guiriec}, {Hadasch},
  {Harding}, {Hays}, {Horan}, {Hughes}, {J{\'o}hannesson}, {Johnson}, {Kamae},
  {Katagiri}, {Kataoka}, {Kawai}, {Kippen}, {Kn{\"o}dlseder}, {Kocevski},
  {Kouveliotou}, {Kuss}, {Lande}, {Latronico}, {Lemoine-Goumard}, {Llena
  Garde}, {Longo}, {Loparco}, {Lott}, {Lovellette}, {Lubrano}, {Makeev},
  {Mazziotta}, {McEnery}, {McGlynn}, {Meegan}, {M{\'e}sz{\'a}ros}, {Michelson},
  {Mitthumsiri}, {Mizuno}, {Moiseev}, {Monte}, {Monzani}, {Moretti},
  {Morselli}, {Moskalenko}, {Murgia}, {Nakajima}, {Nakamori}, {Nolan},
  {Norris}, {Nuss}, {Ohno}, {Ohsugi}, {Omodei}, {Orlando}, {Ormes}, {Ozaki},
  {Paciesas}, {Paneque}, {Panetta}, {Parent}, {Pelassa}, {Pepe},
  {Pesce-Rollins}, {Piron}, {Preece}, {Rain{\`o}}, {Rando}, {Razzano},
  {Razzaque}, {Reimer}, {Ritz}, {Rodriguez}, {Roth}, {Ryde}, {Sadrozinski},
  {Sander}, {Scargle}, {Schalk}, {Sgr{\`o}}, {Siskind}, {Smith}, {Spandre},
  {Spinelli}, {Stamatikos}, {Stecker}, {Strickman}, {Suson}, {Tajima},
  {Takahashi}, {Takahashi}, {Tanaka}, {Thayer}, {Thayer}, {Thompson},
  {Tibaldo}, {Toma}, {Torres}, {Tosti}, {Tramacere}, {Uchiyama}, {Uehara},
  {Usher}, {van der Horst}, {Vasileiou}, {Vilchez}, {Vitale}, {von Kienlin},
  {Waite}, {Wang}, {Wilson-Hodge}, {Winer}, {Wu}, {Yamazaki}, {Yang}, {Ylinen},
  \& {Ziegler}}]{2010ApJ...716.1178A}
{Ackermann}, M., {Asano}, K., {Atwood}, W.~B., {et~al.} 2010, \apj, 716, 1178

\bibitem[{{Ackermann} {et~al.}(2011){Ackermann}, {Ajello}, {Asano}, {Axelsson},
  {Baldini}, {Ballet}, {Barbiellini}, {Baring}, {Bastieri}, {Bechtol},
  {Bellazzini}, {Berenji}, {Bhat}, {Bissaldi}, {Blandford}, {Bonamente},
  {Borgland}, {Bouvier}, {Bregeon}, {Brez}, {Briggs}, {Brigida}, {Bruel},
  {Buehler}, {Buson}, {Caliandro}, {Cameron}, {Caraveo}, {Carrigan},
  {Casandjian}, {Cecchi}, {{\c C}elik}, {Chaplin}, {Charles}, {Chekhtman},
  {Chiang}, {Ciprini}, {Claus}, {Cohen-Tanugi}, {Connaughton}, {Conrad},
  {Cutini}, {Dermer}, {de Angelis}, {de Palma}, {Dingus}, {Silva}, {Drell},
  {Dubois}, {Favuzzi}, {Fegan}, {Ferrara}, {Focke}, {Frailis}, {Fukazawa},
  {Funk}, {Fusco}, {Gargano}, {Gasparrini}, {Gehrels}, {Germani}, {Giglietto},
  {Giordano}, {Giroletti}, {Glanzman}, {Godfrey}, {Goldstein}, {Granot},
  {Greiner}, {Grenier}, {Grove}, {Guiriec}, {Hadasch}, {Hanabata}, {Harding},
  {Hayashi}, {Hayashida}, {Hays}, {Horan}, {Hughes}, {Itoh}, {J{\'o}hannesson},
  {Johnson}, {Johnson}, {Kamae}, {Katagiri}, {Kataoka}, {Kippen},
  {Kn{\"o}dlseder}, {Kocevski}, {Kouveliotou}, {Kuss}, {Lande}, {Latronico},
  {Lee}, {Llena Garde}, {Longo}, {Loparco}, {Lovellette}, {Lubrano}, {Makeev},
  {Mazziotta}, {McBreen}, {McEnery}, {McGlynn}, {Meegan}, {Mehault},
  {M{\'e}sz{\'a}ros}, {Michelson}, {Mizuno}, {Monte}, {Monzani}, {Moretti},
  {Morselli}, {Moskalenko}, {Murgia}, {Nakajima}, {Nakamori}, {Naumann-Godo},
  {Nishino}, {Nolan}, {Norris}, {Nuss}, {Ohno}, {Ohsugi}, {Okumura}, {Omodei},
  {Orlando}, {Ormes}, {Ozaki}, {Paciesas}, {Paneque}, {Panetta}, {Parent},
  {Pelassa}, {Pepe}, {Pesce-Rollins}, {Petrosian}, {Piron}, {Porter}, {Preece},
  {Racusin}, {Rain{\`o}}, {Rando}, {Rau}, {Razzano}, {Razzaque}, {Reimer},
  {Reimer}, {Reposeur}, {Reyes}, {Ripken}, {Ritz}, {Roth}, {Ryde},
  {Sadrozinski}, {Sander}, {Scargle}, {Schalk}, {Sgr{\`o}}, {Siskind}, {Smith},
  {Spandre}, {Spinelli}, {Stamatikos}, {Stecker}, {Strickman}, {Suson},
  {Tajima}, {Takahashi}, {Tanaka}, {Tanaka}, {Thayer}, {Thayer}, {Tibaldo},
  {Tierney}, {Toma}, {Torres}, {Tosti}, {Tramacere}, {Uchiyama}, {Uehara},
  {Usher}, {Vandenbroucke}, {van der Horst}, {Vasileiou}, {Vilchez}, {Vitale},
  {von Kienlin}, {Waite}, {Wang}, {Wilson-Hodge}, {Winer}, {Wood}, {Wu},
  {Yamazaki}, {Yang}, {Ylinen}, \& {Ziegler}}]{2011ApJ...729..114A}
{Ackermann}, M., {Ajello}, M., {Asano}, K., {et~al.} 2011, \apj, 729, 114

\bibitem[{Ackermann {et~al.}(2013)Ackermann, Ajello, Asano, Axelsson, Baldini,
  Ballet, Barbiellini, Bastieri, Bechtol, Bellazzini, Bhat, Bissaldi, Bloom,
  Bonamente, Bonnell, Bouvier, Brandt, Bregeon, Brigida, Bruel, Buehler,
  Burgess, Buson, Byrne, Caliandro, Cameron, Caraveo, Cecchi, Charles, Chaves,
  Chekhtman, Chiang, Chiaro, Ciprini, Claus, Cohen-Tanugi, Connaughton, Conrad,
  Cutini, D'Ammando, de~Angelis, de~Palma, Dermer, Desiante, Digel, Dingus,
  Venere, Drell, Drlica-Wagner, Dubois, Favuzzi, Ferrara, Fitzpatrick, Foley,
  Franckowiak, Fukazawa, Fusco, Gargano, Gasparrini, Gehrels, Germani,
  Giglietto, Giommi, Giordano, Giroletti, Glanzman, Godfrey, Goldstein, Granot,
  Grenier, Grove, Gruber, Guiriec, Hadasch, Hanabata, Hayashida, Horan, Hou,
  Hughes, Inoue, Jackson, Jogler, Jóhannesson, Johnson, Johnson, Kamae,
  Kataoka, Kawano, Kippen, Knödlseder, Kocevski, Kouveliotou, Kuss, Lande,
  Larsson, Latronico, Lee, Longo, Loparco, Lovellette, Lubrano, Massaro, Mayer,
  Mazziotta, McBreen, McEnery, McGlynn, Michelson, Mizuno, Moiseev, Monte,
  Monzani, Moretti, Morselli, Murgia, Nemmen, Nuss, Nymark, Ohno, Ohsugi,
  Omodei, Orienti, Orlando, Paciesas, Paneque, Panetta, Pelassa, Perkins,
  Pesce-Rollins, Piron, Pivato, Porter, Preece, Racusin, Rainò, Rando, Rau,
  Razzano, Razzaque, Reimer, Reimer, Reposeur, Ritz, Romoli, Roth, Ryde,
  Parkinson, Schalk, Sgrò, Siskind, Sonbas, Spandre, Spinelli, Suson, Tajima,
  Takahashi, Takeuchi, Tanaka, Thayer, Thayer, Thompson, Tibaldo, Tierney,
  Tinivella, Torres, Tosti, Troja, Tronconi, Usher, Vandenbroucke, van~der
  Horst, Vasileiou, Vianello, Vitale, von Kienlin, Winer, Wood, Wood, Xiong, \&
  Yang}]{0067-0049-209-1-11}
Ackermann, M., Ajello, M., Asano, K., {et~al.} 2013, The Astrophysical Journal
  Supplement Series, 209, 11

\bibitem[{Ackermann {et~al.}(2014)Ackermann, Ajello, Asano, Atwood, Axelsson,
  Baldini, Ballet, Barbiellini, Baring, Bastieri, Bechtol, Bellazzini,
  Bissaldi, Bonamente, Bregeon, Brigida, Bruel, Buehler, Burgess, Buson,
  Caliandro, Cameron, Caraveo, Cecchi, Chaplin, Charles, Chekhtman, Cheung,
  Chiang, Chiaro, Ciprini, Claus, Cleveland, Cohen-Tanugi, Collazzi, Cominsky,
  Connaughton, Conrad, Cutini, D’Ammando, de~Angelis, DeKlotz, de~Palma,
  Dermer, Desiante, Diekmann, Di~Venere, Drell, Drlica-Wagner, Favuzzi, Fegan,
  Ferrara, Finke, Fitzpatrick, Focke, Franckowiak, Fukazawa, Funk, Fusco,
  Gargano, Gehrels, Germani, Gibby, Giglietto, Giles, Giordano, Giroletti,
  Godfrey, Granot, Grenier, Grove, Gruber, Guiriec, Hadasch, Hanabata, Harding,
  Hayashida, Hays, Horan, Hughes, Inoue, Jogler, Jóhannesson, Johnson, Kawano,
  Knödlseder, Kocevski, Kuss, Lande, Larsson, Latronico, Longo, Loparco,
  Lovellette, Lubrano, Mayer, Mazziotta, McEnery, Michelson, Mizuno, Moiseev,
  Monzani, Moretti, Morselli, Moskalenko, Murgia, Nemmen, Nuss, Ohno, Ohsugi,
  Okumura, Omodei, Orienti, Paneque, Pelassa, Perkins, Pesce-Rollins,
  Petrosian, Piron, Pivato, Porter, Racusin, Rainò, Rando, Razzano, Razzaque,
  Reimer, Reimer, Ritz, Roth, Ryde, Sartori, Parkinson, Scargle, Schulz, Sgrò,
  Siskind, Sonbas, Spandre, Spinelli, Tajima, Takahashi, Thayer, Thayer,
  Thompson, Tibaldo, Tinivella, Torres, Tosti, Troja, Usher, Vandenbroucke,
  Vasileiou, Vianello, Vitale, Winer, Wood, Yamazaki, Younes, Yu, Zhu, Bhat,
  Briggs, Byrne, Foley, Goldstein, Jenke, Kippen, Kouveliotou, McBreen, Meegan,
  Paciesas, Preece, Rau, Tierney, van~der Horst, von Kienlin, Wilson-Hodge,
  Xiong, Cusumano, La~Parola, \& Cummings}]{Ackermann03012014}
---. 2014, Science, 343, 42

\bibitem[{{Actis} {et~al.}(2011){Actis}, {Agnetta}, {Aharonian},
  {Akhperjanian}, {Aleksi{\'c}}, {Aliu}, {Allan}, {Allekotte}, {Antico},
  {Antonelli}, \& et~al.}]{2011ExA....32..193A}
{Actis}, M., {Agnetta}, G., {Aharonian}, F., {et~al.} 2011, Experimental
  Astronomy, 32, 193

\bibitem[{{Aharonian} {et~al.}(2006){Aharonian}, {Akhperjanian}, {Bazer-Bachi},
  {Beilicke}, {Benbow}, {Berge}, {Bernl{\"o}hr}, {Boisson}, {Bolz}, {Borrel},
  {Braun}, {Breitling}, {Brown}, {Chadwick}, {Chounet}, {Cornils},
  {Costamante}, {Degrange}, {Dickinson}, {Djannati-Ata{\"i}}, {Drury}, {Dubus},
  {Emmanoulopoulos}, {Espigat}, {Feinstein}, {Fontaine}, {Fuchs}, {Funk},
  {Gallant}, {Giebels}, {Gillessen}, {Glicenstein}, {Goret}, {Hadjichristidis},
  {Hauser}, {Heinzelmann}, {Henri}, {Hermann}, {Hinton}, {Hofmann}, {Holleran},
  {Horns}, {Jacholkowska}, {de Jager}, {Kh{\'e}lifi}, {Komin}, {Konopelko},
  {Latham}, {Le Gallou}, {Lemi{\`e}re}, {Lemoine-Goumard}, {Leroy}, {Lohse},
  {Martin}, {Martineau-Huynh}, {Marcowith}, {Masterson}, {McComb}, {de
  Naurois}, {Nolan}, {Noutsos}, {Orford}, {Osborne}, {Ouchrif}, {Panter},
  {Pelletier}, {Pita}, {P{\"u}hlhofer}, {Punch}, {Raubenheimer}, {Raue},
  {Raux}, {Rayner}, {Reimer}, {Reimer}, {Ripken}, {Rob}, {Rolland}, {Rowell},
  {Sahakian}, {Saug{\'e}}, {Schlenker}, {Schlickeiser}, {Schuster}, {Schwanke},
  {Siewert}, {Sol}, {Spangler}, {Steenkamp}, {Stegmann}, {Tavernet}, {Terrier},
  {Th{\'e}oret}, {Tluczykont}, {Vasileiadis}, {Venter}, {Vincent}, {V{\"o}lk},
  \& {Wagner}}]{2006ApJ...636..777A}
{Aharonian}, F., {Akhperjanian}, A.~G., {Bazer-Bachi}, A.~R., {et~al.} 2006,
  \apj, 636, 777

\bibitem[{{Albert} {et~al.}(2006){Albert}, {Aliu}, {Anderhub}, {Antoranz},
  {Armada}, {Asensio}, {Baixeras}, {Barrio}, {Bartelt}, {Bartko}, {Bastieri},
  {Bavikadi}, {Bednarek}, {Berger}, {Bigongiari}, {Biland}, {Bisesi}, {Bock},
  {Bretz}, {Britvitch}, {Camara}, {Chilingarian}, {Ciprini}, {Coarasa},
  {Commichau}, {Contreras}, {Cortina}, {Curtef}, {Danielyan}, {Dazzi}, {De
  Angelis}, {de los Reyes}, {De Lotto}, {Domingo-Santamar{\'{\i}}a}, {Dorner},
  {Doro}, {Errando}, {Fagiolini}, {Ferenc}, {Fern{\'a}ndez}, {Firpo}, {Flix},
  {Fonseca}, {Font}, {Galante}, {Garczarczyk}, {Gaug}, {Giller}, {Goebel},
  {Hakobyan}, {Hayashida}, {Hengstebeck}, {H{\"o}hne}, {Hose}, {Jaco{\'n}},
  {Kalekin}, {Kranich}, {Laille}, {Lenisa}, {Liebing}, {Lindfors}, {Longo},
  {L{\'o}pez}, {L{\'o}pez}, {Lorenz}, {Lucarelli}, {Majumdar}, {Maneva},
  {Mannheim}, {Mariotti}, {Mart{\'{\i}}nez}, {Mase}, {Mazin}, {Meucci},
  {Meyer}, {Miranda}, {Mirzoyan}, {Mizobuchi}, {Moralejo}, {Nilsson},
  {O{\~n}a-Wilhelmi}, {Ordu{\~n}a}, {Otte}, {Oya}, {Paneque}, {Paoletti},
  {Pasanen}, {Pascoli}, {Pauss}, {Pavel}, {Pegna}, {Persic}, {Peruzzo},
  {Piccioli}, {Prandini}, {Rhode}, {Rico}, {Riegel}, {Rissi}, {Robert},
  {R{\"u}gamer}, {Saggion}, {S{\'a}nchez}, {Sartori}, {Scalzotto}, {Schmitt},
  {Schweizer}, {Shayduk}, {Shinozaki}, {Shore}, {Sidro}, {Sillanp{\"a}{\"a}},
  {Sobczy{\'n}ska}, {Stamerra}, {Stark}, {Takalo}, {Temnikov}, {Tescaro},
  {Teshima}, {Tonello}, {Torres}, {Torres}, {Turini}, {Vankov}, {Vardanyan},
  {Vitale}, {Wagner}, {Wibig}, {Wittek}, \& {Zapatero}}]{Albert+06magic050713a}
{Albert}, J., {Aliu}, E., {Anderhub}, H., {et~al.} 2006, \apjl, 641, L9

\bibitem[{{Ando} {et~al.}(2012){Ando}, {Baret}, {Bouhou}, {Chassande-Mottin},
  {Kouchner}, {Moscoso}, {Van Elewyck}, {Bartos}, {M{\'a}rka}, {M{\'a}rka},
  {Corsi}, {Di Palma}, {Papa}, {Dietz}, {Donzaud}, {Eichler}, {Finley},
  {Guetta}, {Halzen}, {Jones}, {Sutton}, {Kandhasamy}, {Mandic}, {Thrane},
  {Kotake}, {Piran}, {Pradier}, {Romero}, \& {Waxman}}]{2012arXiv1203.5192A}
{Ando}, S., {Baret}, B., {Bouhou}, B., {et~al.} 2012, ArXiv 1203.5192

\bibitem[{{Atwood} {et~al.}(2009){Atwood}, {Abdo}, {Ackermann}, {Althouse},
  {Anderson}, {Axelsson}, {Baldini}, {Ballet}, {Band}, {Barbiellini}, \&
  et~al.}]{2009ApJ...697.1071A}
{Atwood}, W.~B., {Abdo}, A.~A., {Ackermann}, M., {et~al.} 2009, \apj, 697, 1071

\bibitem[{{Barthelmy} {et~al.}(2005){Barthelmy}, {Barbier}, {Cummings},
  {Fenimore}, {Gehrels}, {Hullinger}, {Krimm}, {Markwardt}, {Palmer},
  {Parsons}, {Sato}, {Suzuki}, {Takahashi}, {Tashiro}, \&
  {Tueller}}]{2005SSRv..120..143B}
{Barthelmy}, S.~D., {Barbier}, L.~M., {Cummings}, J.~R., {et~al.} 2005, \ssr,
  120, 143

\bibitem[{{Bartos} {et~al.}(2013){Bartos}, {Brady}, \&
  {M{\'a}rka}}]{2013CQGra..30l3001B}
{Bartos}, I., {Brady}, P., \& {M{\'a}rka}, S. 2013, Classical and Quantum
  Gravity, 30, 123001

\bibitem[{Bartos {et~al.}(2011)Bartos, Finley, Corsi, \&
  M\'arka}]{PhysRevLett.107.251101}
Bartos, I., Finley, C., Corsi, A., \& M\'arka, S. 2011, Phys. Rev. Lett., 107,
  251101

\bibitem[{{Beloborodov}(2010)}]{2010MNRAS.407.1033B}
{Beloborodov}, A.~M. 2010, \mnras, 407, 1033

\bibitem[{{Berger}(2013)}]{2013arXiv1311.2603B}
{Berger}, E. 2013, ArXiv 1311.2603

\bibitem[{{Bernl{\"o}hr} {et~al.}(2013){Bernl{\"o}hr}, {Barnacka}, {Becherini},
  {Blanch Bigas}, {Carmona}, {Colin}, {Decerprit}, {Di Pierro}, {Dubois},
  {Farnier}, {Funk}, {Hermann}, {Hinton}, {Humensky}, {Kh{\'e}lifi}, {Kihm},
  {Komin}, {Lenain}, {Maier}, {Mazin}, {Medina}, {Moralejo}, {Nolan}, {Ohm},
  {de O{\~n}a Wilhelmi}, {Parsons}, {Paz Arribas}, {Pedaletti}, {Pita},
  {Prokoph}, {Rulten}, {Schwanke}, {Shayduk}, {Stamatescu}, {Vallania},
  {Vorobiov}, {Wischnewski}, {Yoshikoshi}, {Zech}, \& {CTA
  Consortium}}]{2013APh....43..171B}
{Bernl{\"o}hr}, K., {Barnacka}, A., {Becherini}, Y., {et~al.} 2013,
  Astroparticle Physics, 43, 171

\bibitem[{{Blandford} \& {McKee}(1976)}]{Blandford+76bm}
{Blandford}, R.~D., \& {McKee}, C.~F. 1976, Physics of Fluids, 19, 1130

\bibitem[{{Blinnikov} {et~al.}(1984){Blinnikov}, {Novikov}, {Perevodchikova},
  \& {Polnarev}}]{1984SvAL...10..177B}
{Blinnikov}, S.~I., {Novikov}, I.~D., {Perevodchikova}, T.~V., \& {Polnarev},
  A.~G. 1984, Soviet Astronomy Letters, 10, 177

\bibitem[{{Bloom} {et~al.}(2009){Bloom}, {Holz}, {Hughes}, {Menou}, {Adams},
  {Anderson}, {Becker}, {Bower}, {Brandt}, {Cobb}, {Cook}, {Corsi}, {Covino},
  {Fox}, {Fruchter}, {Fryer}, {Grindlay}, {Hartmann}, {Haiman}, {Kocsis},
  {Jones}, {Loeb}, {Marka}, {Metzger}, {Nakar}, {Nissanke}, {Perley}, {Piran},
  {Poznanski}, {Prince}, {Schnittman}, {Soderberg}, {Strauss}, {Shawhan},
  {Shoemaker}, {Sievers}, {Stubbs}, {Tagliaferri}, {Ubertini}, \&
  {Wozniak}}]{2009arXiv0902.1527B}
{Bloom}, J.~S., {Holz}, D.~E., {Hughes}, S.~A., {et~al.} 2009, ArXiv 0902.1527

\bibitem[{{Bouvier} {et~al.}(2011){Bouvier}, {Gilmore}, {Connaughton}, {Otte},
  {Primack}, \& {Williams}}]{2011arXiv1109.5680B}
{Bouvier}, A., {Gilmore}, R., {Connaughton}, V., {et~al.} 2011, ArXiv 1109.5680

\bibitem[{{Braga} {et~al.}(2004){Braga}, {Rothschild}, {Heise}, {Staubert},
  {Remillard}, {D'Amico}, {Jablonski}, {Heindl}, {Matteson}, {Kuulkers},
  {Wilms}, \& {Kendziorra}}]{2004AdSpR..34.2657B}
{Braga}, J., {Rothschild}, R., {Heise}, J., {et~al.} 2004, Advances in Space
  Research, 34, 2657

\bibitem[{{Bromberg} {et~al.}(2013){Bromberg}, {Nakar}, {Piran}, \&
  {Sari}}]{2013ApJ...764..179B}
{Bromberg}, O., {Nakar}, E., {Piran}, T., \& {Sari}, R. 2013, \apj, 764, 179

\bibitem[{{Cannon} {et~al.}(2012){Cannon}, {Cariou}, {Chapman},
  {Crispin-Ortuzar}, {Fotopoulos}, {Frei}, {Hanna}, {Kara}, {Keppel}, {Liao},
  {Privitera}, {Searle}, {Singer}, \& {Weinstein}}]{2012ApJ...748..136C}
{Cannon}, K., {Cariou}, R., {Chapman}, A., {et~al.} 2012, \apj, 748, 136

\bibitem[{{Cavalier} {et~al.}(2006){Cavalier}, {Barsuglia}, {Bizouard},
  {Brisson}, {Clapson}, {Davier}, {Hello}, {Kreckelbergh}, {Leroy}, \&
  {Varvella}}]{2006PhRvD..74h2004C}
{Cavalier}, F., {Barsuglia}, M., {Bizouard}, M.-A., {et~al.} 2006, \prd, 74,
  082004

\bibitem[{{CHEN}(2011)}]{2011ICRC....8..240C}
{CHEN}, P. 2011, International Cosmic Ray Conference, 8, 240

\bibitem[{Connaughton(2014)}]{Valerie}
Connaughton, V., e. 2014, submitted to ApJS

\bibitem[{{Coward} {et~al.}(2012){Coward}, {Howell}, {Piran}, {Stratta},
  {Branchesi}, {Bromberg}, {Gendre}, {Burman}, \&
  {Guetta}}]{2012MNRAS.425.2668C}
{Coward}, D.~M., {Howell}, E.~J., {Piran}, T., {et~al.} 2012, \mnras, 425, 2668

\bibitem[{{de Jager} {et~al.}(1996){de Jager}, {Harding}, {Michelson}, {Nel},
  {Nolan}, {Sreekumar}, \& {Thompson}}]{dejager+96maxsyn}
{de Jager}, O.~C., {Harding}, A.~K., {Michelson}, P.~F., {et~al.} 1996, \apj,
  457, 253

\bibitem[{{De Pasquale} {et~al.}(2010){De Pasquale}, {Schady}, {Kuin}, {Page},
  {Curran}, {Zane}, {Oates}, {Holland}, {Breeveld}, {Hoversten}, {Chincarini},
  {Grupe}, {Abdo}, {Ackermann}, {Ajello}, {Axelsson}, {Baldini}, {Ballet},
  {Barbiellini}, {Baring}, {Bastieri}, {Bechtol}, {Bellazzini}, {Berenji},
  {Bissaldi}, {Blandford}, {Bloom}, {Bonamente}, {Borgland}, {Bouvier},
  {Bregeon}, {Brez}, {Briggs}, {Brigida}, {Bruel}, {Burnett}, {Buson},
  {Caliandro}, {Cameron}, {Caraveo}, {Carrigan}, {Casandjian}, {Cecchi}, {{\c
  C}elik}, {Chekhtman}, {Chiang}, {Ciprini}, {Claus}, {Cohen-Tanugi},
  {Connaughton}, {Conrad}, {Dermer}, {de Angelis}, {de Palma}, {Dingus},
  {Silva}, {Drell}, {Dubois}, {Dumora}, {Farnier}, {Favuzzi}, {Fegan},
  {Fishman}, {Focke}, {Frailis}, {Fukazawa}, {Funk}, {Fusco}, {Gargano},
  {Gasparrini}, {Gehrels}, {Germani}, {Giglietto}, {Giordano}, {Glanzman},
  {Godfrey}, {Granot}, {Greiner}, {Grenier}, {Grove}, {Guillemot}, {Guiriec},
  {Harding}, {Hayashida}, {Hays}, {Horan}, {Hughes}, {Jackson},
  {J{\'o}hannesson}, {Johnson}, {Johnson}, {Kamae}, {Katagiri}, {Kataoka},
  {Kawai}, {Kerr}, {Kippen}, {Kn{\"o}dlseder}, {Kocevski}, {Kuss}, {Lande},
  {Latronico}, {Lemoine-Goumard}, {Longo}, {Loparco}, {Lott}, {Lovellette},
  {Lubrano}, {Makeev}, {Mazziotta}, {McEnery}, {McGlynn}, {Meegan},
  {M{\'e}sz{\'a}ros}, {Meurer}, {Michelson}, {Mitthumsiri}, {Mizuno}, {Monte},
  {Monzani}, {Moretti}, {Morselli}, {Moskalenko}, {Murgia}, {Nolan}, {Norris},
  {Nuss}, {Ohno}, {Ohsugi}, {Omodei}, {Orlando}, {Ormes}, {Paciesas},
  {Paneque}, {Panetta}, {Parent}, {Pelassa}, {Pepe}, {Pesce-Rollins}, {Piron},
  {Porter}, {Preece}, {Rain{\`o}}, {Rando}, {Razzano}, {Reimer}, {Reimer},
  {Reposeur}, {Ritz}, {Rochester}, {Rodriguez}, {Roth}, {Ryde}, {Sadrozinski},
  {Sander}, {Saz Parkinson}, {Scargle}, {Schalk}, {Sgr{\`o}}, {Siskind},
  {Smith}, {Spandre}, {Spinelli}, {Stamatikos}, {Starck}, {Stecker},
  {Strickman}, {Suson}, {Tajima}, {Takahashi}, {Tanaka}, {Thayer}, {Thayer},
  {Thompson}, {Tibaldo}, {Toma}, {Torres}, {Tosti}, {Tramacere}, {Uchiyama},
  {Uehara}, {Usher}, {van der Horst}, {Vasileiou}, {Vilchez}, {Vitale}, {von
  Kienlin}, {Waite}, {Wang}, {Winer}, {Wood}, {Wu}, {Yamazaki}, {Ylinen}, \&
  {Ziegler}}]{2010ApJ...709L.146D}
{De Pasquale}, M., {Schady}, P., {Kuin}, N.~P.~M., {et~al.} 2010, \apjl, 709,
  L146

\bibitem[{{de Vos} {et~al.}(2009){de Vos}, {Gunst}, \&
  {Nijboer}}]{2009IEEEP..97.1431D}
{de Vos}, M., {Gunst}, A.~W., \& {Nijboer}, R. 2009, IEEE Proceedings, 97, 1431

\bibitem[{{DeYoung} \& {HAWC Collaboration}(2012)}]{2012NIMPA.692...72D}
{DeYoung}, T., \& {HAWC Collaboration}. 2012, Nuclear Instruments and Methods
  in Physics Research A, 692, 72

\bibitem[{{Dom{\'{\i}}nguez} {et~al.}(2011){Dom{\'{\i}}nguez}, {Primack},
  {Rosario}, {Prada}, {Gilmore}, {Faber}, {Koo}, {Somerville},
  {P{\'e}rez-Torres}, {P{\'e}rez-Gonz{\'a}lez}, {Huang}, {Davis},
  {Guhathakurta}, {Barmby}, {Conselice}, {Lozano}, {Newman}, \&
  {Cooper}}]{2011MNRAS.410.2556D}
{Dom{\'{\i}}nguez}, A., {Primack}, J.~R., {Rosario}, D.~J., {et~al.} 2011,
  \mnras, 410, 2556

\bibitem[{{Doro} {et~al.}(2013){Doro}, {Conrad}, {Emmanoulopoulos},
  {S{\`a}nchez-Conde}, {Barrio}, {Birsin}, {Bolmont}, {Brun}, {Colafrancesco},
  {Connell}, {Contreras}, {Daniel}, {Fornasa}, {Gaug}, {Glicenstein},
  {Gonz{\'a}lez-Mu{\~n}oz}, {Hassan}, {Horns}, {Jacholkowska}, {Jahn},
  {Mazini}, {Mirabal}, {Moralejo}, {Moulin}, {Nieto}, {Ripken}, {Sandaker},
  {Schwanke}, {Spengler}, {Stamerra}, {Viana}, {Zechlin}, {Zimmer}, \& {CTA
  Consortium}}]{2013APh....43..189D}
{Doro}, M., {Conrad}, J., {Emmanoulopoulos}, D., {et~al.} 2013, Astroparticle
  Physics, 43, 189

\bibitem[{{Dubus} {et~al.}(2013){Dubus}, {Contreras}, {Funk}, {Gallant},
  {Hassan}, {Hinton}, {Inoue}, {Kn{\"o}dlseder}, {Martin}, {Mirabal}, {de
  Naurois}, {Renaud}, \& {CTA Consortium}}]{2013APh....43..317D}
{Dubus}, G., {Contreras}, J.~L., {Funk}, S., {et~al.} 2013, Astroparticle
  Physics, 43, 317

\bibitem[{{Eichler} {et~al.}(1989){Eichler}, {Livio}, {Piran}, \&
  {Schramm}}]{1989Natur.340..126E}
{Eichler}, D., {Livio}, M., {Piran}, T., \& {Schramm}, D.~N. 1989, \nat, 340,
  126

\bibitem[{{Ekers}(2003)}]{2003ASPC..289...21E}
{Ekers}, R.~D. 2003, in Astronomical Society of the Pacific Conference Series,
  Vol. 289, The Proceedings of the IAU 8th Asian-Pacific Regional Meeting,
  Volume 1, ed. S.~{Ikeuchi}, J.~{Hearnshaw}, \& T.~{Hanawa}, 21--28

\bibitem[{{Evans} {et~al.}(2012){Evans}, {Fridriksson}, {Gehrels}, {Homan},
  {Osborne}, {Siegel}, {Beardmore}, {Handbauer}, {Gelbord}, {Kennea}, \&
  et~al.}]{2012ApJS..203...28E}
{Evans}, P.~A., {Fridriksson}, J.~K., {Gehrels}, N., {et~al.} 2012, \apjs, 203,
  28

\bibitem[{{Fairhurst}(2009)}]{2009NJPh...11l3006F}
{Fairhurst}, S. 2009, New Journal of Physics, 11, 123006

\bibitem[{{Fairhurst}(2011)}]{2011CQGra..28j5021F}
---. 2011, Classical and Quantum Gravity, 28, 105021

\bibitem[{{Fern{\'a}ndez} \& {Metzger}(2013)}]{2013MNRAS.435..502F}
{Fern{\'a}ndez}, R., \& {Metzger}, B.~D. 2013, \mnras, 435, 502

\bibitem[{{Fong} \& {Berger}(2013)}]{2013ApJ...776...18F}
{Fong}, W., \& {Berger}, E. 2013, \apj, 776, 18

\bibitem[{{Ghisellini} {et~al.}(2010){Ghisellini}, {Ghirlanda}, {Nava}, \&
  {Celotti}}]{2010MNRAS.403..926G}
{Ghisellini}, G., {Ghirlanda}, G., {Nava}, L., \& {Celotti}, A. 2010, \mnras,
  403, 926

\bibitem[{{G{\"o}tz} {et~al.}(2009){G{\"o}tz}, {Paul}, {Basa}, {Wei}, {Zhang},
  {Atteia}, {Barret}, {Cordier}, {Claret}, {Deng}, {Fan}, {Hu}, {Huang},
  {Mandrou}, {Mereghetti}, {Qiu}, \& {Wu}}]{2009AIPC.1133...25G}
{G{\"o}tz}, D., {Paul}, J., {Basa}, S., {et~al.} 2009, in American Institute of
  Physics Conference Series, Vol. 1133, American Institute of Physics
  Conference Series, ed. C.~{Meegan}, C.~{Kouveliotou}, \& N.~{Gehrels}, 25--30

\bibitem[{{Gou} \& {M{\'e}sz{\'a}ros}(2007)}]{Gou+07glast}
{Gou}, L.-J., \& {M{\'e}sz{\'a}ros}, P. 2007, \apj, 668, 392

\bibitem[{{Granot} \& {Sari}(2002)}]{Granot+02Dabreaks}
{Granot}, J., \& {Sari}, R. 2002, \apj, 568, 820

\bibitem[{{Guetta} \& {Piran}(2006)}]{2006A&A...453..823G}
{Guetta}, D., \& {Piran}, T. 2006, \aap, 453, 823

\bibitem[{{Hasco{\"e}t} {et~al.}(2012){Hasco{\"e}t}, {Daigne}, {Mochkovitch},
  \& {Vennin}}]{2012MNRAS.421..525H}
{Hasco{\"e}t}, R., {Daigne}, F., {Mochkovitch}, R., \& {Vennin}, V. 2012,
  \mnras, 421, 525

\bibitem[{{He} {et~al.}(2012){He}, {Liu}, {Wang}, {Nagataki}, {Murase}, \&
  {Dai}}]{2012ApJ...752...29H}
{He}, H.-N., {Liu}, R.-Y., {Wang}, X.-Y., {et~al.} 2012, \apj, 752, 29

\bibitem[{Hillas(2013)}]{Hillas201319}
Hillas, A. 2013, Astroparticle Physics, 43, 19 , seeing the High-Energy
  Universe with the Cherenkov Telescope Array - The Science Explored with the
  \{CTA\}

\bibitem[{H\"ummer {et~al.}(2012)H\"ummer, Baerwald, \&
  Winter}]{PhysRevLett.108.231101}
H\"ummer, S., Baerwald, P., \& Winter, W. 2012, Phys. Rev. Lett., 108, 231101

\bibitem[{{Hurley} {et~al.}(2011){Hurley}, {Golenetskii}, {Aptekar}, {Mazets},
  {Pal'Shin}, {Frederiks}, {Mitrofanov}, {Golovin}, {Kozyrev}, {Litvak},
  {Sanin}, {Boynton}, {Fellows}, {Harshman}, {Starr}, {von Kienlin}, {Rau},
  {Yamaoka}, {Ohno}, {Fukazawa}, {Takahashi}, {Tashiro}, {Terada}, {Murakami},
  {Makishima}, {Barthelmy}, {Cummings}, {Gehrels}, {Krimm}, {Cline},
  {Goldsten}, {Del Monte}, {Feroci}, {Marisaldi}, {Briggs}, {Connaughton},
  {Meegan}, {Smith}, {Wigger}, \& {Hajdas}}]{2011AIPC.1358..385H}
{Hurley}, K., {Golenetskii}, S., {Aptekar}, R., {et~al.} 2011, in American
  Institute of Physics Conference Series, Vol. 1358, American Institute of
  Physics Conference Series, ed. J.~E. {McEnery}, J.~L. {Racusin}, \&
  N.~{Gehrels}, 385--388

\bibitem[{{Inoue} {et~al.}(2013){Inoue}, {Granot}, {O'Brien}, {Asano},
  {Bouvier}, {Carosi}, {Connaughton}, {Garczarczyk}, {Gilmore}, {Hinton},
  {Inoue}, {Ioka}, {Kakuwa}, {Markoff}, {Murase}, {Osborne}, {Otte},
  {Starling}, {Tajima}, {Teshima}, {Toma}, {Wagner}, {Wijers}, {Williams},
  {Yamamoto}, {Yamazaki}, \& {CTA Consortium}}]{Inoue+13GRBCTA}
{Inoue}, S., {Granot}, J., {O'Brien}, P.~T., {et~al.} 2013, Astroparticle
  Physics, 43, 252

\bibitem[{{Kanner} {et~al.}(2012){Kanner}, {Camp}, {Racusin}, {Gehrels}, \&
  {White}}]{2012ApJ...759...22K}
{Kanner}, J., {Camp}, J., {Racusin}, J., {Gehrels}, N., \& {White}, D. 2012,
  \apj, 759, 22

\bibitem[{{Kanner} {et~al.}(2008){Kanner}, {Huard}, {M{\'a}rka}, {Murphy},
  {Piscionere}, {Reed}, \& {Shawhan}}]{2008CQGra..25r4034K}
{Kanner}, J., {Huard}, T.~L., {M{\'a}rka}, S., {et~al.} 2008, Classical and
  Quantum Gravity, 25, 184034

\bibitem[{{Kasen} {et~al.}(2013){Kasen}, {Badnell}, \&
  {Barnes}}]{2013ApJ...774...25K}
{Kasen}, D., {Badnell}, N.~R., \& {Barnes}, J. 2013, \apj, 774, 25

\bibitem[{{Kasliwal} \& {Nissanke}(2013)}]{2013arXiv1309.1554K}
{Kasliwal}, M.~M., \& {Nissanke}, S. 2013, ArXiv e-prints

\bibitem[{{Klimenko} {et~al.}(2011){Klimenko}, {Vedovato}, {Drago}, {Mazzolo},
  {Mitselmakher}, {Pankow}, {Prodi}, {Re}, {Salemi}, \&
  {Yakushin}}]{2011PhRvD..83j2001K}
{Klimenko}, S., {Vedovato}, G., {Drago}, M., {et~al.} 2011, \prd, 83, 102001

\bibitem[{{Kulkarni}(2005)}]{2005astro.ph.10256K}
{Kulkarni}, S.~R. 2005, ArXiv 0510256

\bibitem[{{Kumar} \& {Barniol Duran}(2010)}]{2010MNRAS.409..226K}
{Kumar}, P., \& {Barniol Duran}, R. 2010, \mnras, 409, 226

\bibitem[{{Kuroda}(2011)}]{2011IJMPD..20.1755K}
{Kuroda}, K. 2011, International Journal of Modern Physics D, 20, 1755

\bibitem[{{Lee} \& {Ramirez-Ruiz}(2007)}]{2007NJPh....9...17L}
{Lee}, W.~H., \& {Ramirez-Ruiz}, E. 2007, New Journal of Physics, 9, 17

\bibitem[{{Li} \& {Paczy{\'n}ski}(1998)}]{1998ApJ...507L..59L}
{Li}, L.-X., \& {Paczy{\'n}ski}, B. 1998, Astrophys. J., 507, L59

\bibitem[{{LIGO} \& {Virgo}(2012{\natexlab{a}})}]{LVCcommissioning}
{LIGO}, \& {Virgo}. 2012{\natexlab{a}}, LIGO-P1200087

\bibitem[{{LIGO} \& {Virgo}(2012{\natexlab{b}})}]{2012A&A...539A.124L}
---. 2012{\natexlab{b}}, \aap, 539, A124

\bibitem[{{LIGO} \& {Virgo}(2013)}]{2013arXiv1304.0670L}
---. 2013, ArXiv 1304.0670

\bibitem[{{Liu} {et~al.}(2013){Liu}, {Wang}, \& {Wu}}]{Liu+13ic}
{Liu}, R.-Y., {Wang}, X.-Y., \& {Wu}, X.-F. 2013, ArXiv 1306.5207

\bibitem[{{Meegan} {et~al.}(2009){Meegan}, {Lichti}, {Bhat}, {Bissaldi},
  {Briggs}, {Connaughton}, {Diehl}, {Fishman}, {Greiner}, {Hoover}, {van der
  Horst}, {von Kienlin}, {Kippen}, {Kouveliotou}, {McBreen}, {Paciesas},
  {Preece}, {Steinle}, {Wallace}, {Wilson}, \&
  {Wilson-Hodge}}]{2009ApJ...702..791M}
{Meegan}, C., {Lichti}, G., {Bhat}, P.~N., {et~al.} 2009, \apj, 702, 791

\bibitem[{{M{\'e}sz{\'a}ros}(2013)}]{2013APh....43..134M}
{M{\'e}sz{\'a}ros}, P. 2013, Astroparticle Physics, 43, 134

\bibitem[{{M{\'e}sz{\'a}ros} \& {Gehrels}(2012)}]{2012RAA....12.1139M}
{M{\'e}sz{\'a}ros}, P., \& {Gehrels}, N. 2012, Research in Astronomy and
  Astrophysics, 12, 1139

\bibitem[{{Metzger} \& {Berger}(2012)}]{2012ApJ...746...48M}
{Metzger}, B.~D., \& {Berger}, E. 2012, \apj, 746, 48

\bibitem[{{Metzger} {et~al.}(2008){Metzger}, {Piro}, \&
  {Quataert}}]{2008MNRAS.390..781M}
{Metzger}, B.~D., {Piro}, A.~L., \& {Quataert}, E. 2008, \mnras, 390, 781

\bibitem[{{Metzger} {et~al.}(2010){Metzger}, {Martinez-Pinedo}, {Darbha},
  {Quataert}, {Arcones}, {Kasen}, {Thomas}, {Nugent}, {Panov}, \&
  {Zinner}}]{2010MNRAS.406.2650M}
{Metzger}, B.~D., {Martinez-Pinedo}, G., {Darbha}, S., {et~al.} 2010, \mnras,
  406, 2650

\bibitem[{{Murase} {et~al.}(2013){Murase}, {Kashiyama}, \&
  {M{\'e}sz{\'a}ros}}]{2013PhRvL.111m1102M}
{Murase}, K., {Kashiyama}, K., \& {M{\'e}sz{\'a}ros}, P. 2013, Physical Review
  Letters, 111, 131102

\bibitem[{{Nakar}(2007)}]{2007PhR...442..166N}
{Nakar}, E. 2007, \physrep, 442, 166

\bibitem[{{Nakar} {et~al.}(2009){Nakar}, {Ando}, \& {Sari}}]{Nakar+09KN}
{Nakar}, E., {Ando}, S., \& {Sari}, R. 2009, \apj, 703, 675

\bibitem[{{Nakar} {et~al.}(2006){Nakar}, {Gal-Yam}, \&
  {Fox}}]{2006ApJ...650..281N}
{Nakar}, E., {Gal-Yam}, A., \& {Fox}, D.~B. 2006, \apj, 650, 281

\bibitem[{{Nakar} \& {Piran}(2011)}]{2011Natur.478...82N}
{Nakar}, E., \& {Piran}, T. 2011, \nat, 478, 82

\bibitem[{{Nissanke} {et~al.}(2013){Nissanke}, {Kasliwal}, \&
  {Georgieva}}]{2013ApJ...767..124N}
{Nissanke}, S., {Kasliwal}, M., \& {Georgieva}, A. 2013, \apj, 767, 124

\bibitem[{{Nissanke} {et~al.}(2011){Nissanke}, {Sievers}, {Dalal}, \&
  {Holz}}]{2011ApJ...739...99N}
{Nissanke}, S., {Sievers}, J., {Dalal}, N., \& {Holz}, D. 2011, \apj, 739, 99

\bibitem[{{Norris} \& {Bonnell}(2006)}]{norris06}
{Norris}, J.~P., \& {Bonnell}, J.~T. 2006, \apj, 643, 266

\bibitem[{{Paczynski}(1986)}]{1986ApJ...308L..43P}
{Paczynski}, B. 1986, \apjl, 308, L43

\bibitem[{{Park} {et~al.}(2012){Park}, {Ahmad}, {Barrillon}, {Brandt},
  {Budtz-J{\o}rgensen}, {Castro-Tirado}, {Chen}, {Choi}, {Connell},
  {Dagoret-Campagne}, {Eyles}, {Grossan}, {Huang}, {Jeong}, {Jung}, {Kim},
  {Kim}, {Kim}, {Kim}, {Krasnov}, {Lee}, {Lim}, {Linder}, {Liu}, {Lund}, {Min},
  {Na}, {Nam}, {Panasyuk}, {Ripa}, {Reglero}, {Rodrigo}, {Smoot}, {Suh},
  {Svertilov}, {Vedenkin}, {Wang}, \& {Yashin}}]{2012SPIE.8443E..0IP}
{Park}, I.~H., {Ahmad}, S., {Barrillon}, P., {et~al.} 2012, in Society of
  Photo-Optical Instrumentation Engineers (SPIE) Conference Series, Vol. 8443,
  Society of Photo-Optical Instrumentation Engineers (SPIE) Conference Series
  
\bibitem[{{Piran} {et~al.}(2013){Piran}, {Nakar}, \&
  {Rosswog}}]{2013MNRAS.430.2121P}
{Piran}, T., {Nakar}, E., \& {Rosswog}, S. 2013, \mnras, 430, 2121

\bibitem[{{Raymond} {et~al.}(2009){Raymond}, {van der Sluys}, {Mandel},
  {Kalogera}, {R{\"o}ver}, \& {Christensen}}]{2009CQGra..26k4007R}
{Raymond}, V., {van der Sluys}, M.~V., {Mandel}, I., {et~al.} 2009, Classical
  and Quantum Gravity, 26, 114007

\bibitem[{{Rees} \& {M\'esz\'aros}(1998)}]{Rees+98refresh}
{Rees}, M.~J., \& {M\'esz\'aros}, P. 1998, \apjl, 496, L1+

\bibitem[{{R{\"o}ver} {et~al.}(2007){R{\"o}ver}, {Meyer}, \&
  {Christensen}}]{2007PhRvD..75f2004R}
{R{\"o}ver}, C., {Meyer}, R., \& {Christensen}, N. 2007, \prd, 75, 062004

\bibitem[{{Sari} \& {Esin}(2001)}]{Sari+01ic}
{Sari}, R., \& {Esin}, A.~A. 2001, \apj, 548, 787

\bibitem[{{Sari} \& {Piran}(1999)}]{1999ApJ...520..641S}
{Sari}, R., \& {Piran}, T. 1999, \apj, 520, 641

\bibitem[{{Schutz}(2011)}]{2011CQGra..28l5023S}
{Schutz}, B.~F. 2011, Classical and Quantum Gravity, 28, 125023

\bibitem[{{Siellez} {et~al.}(2014){Siellez}, {Bo{\"e}r}, \&
  {Gendre}}]{2014MNRAS.437..649S}
{Siellez}, K., {Bo{\"e}r}, M., \& {Gendre}, B. 2014, \mnras, 437, 649

\bibitem[{{Singer} {et~al.}(2014){Singer}, {Price}, {Farr}, {Urban}, {Pankow},
  {Vitale}, {Veitch}, {Farr}, {Hanna}, {Cannon}, {Downes}, {Graff}, {Haster},
  {Mandel}, {Sidery}, \& {Vecchio}}]{2014arXiv1404.5623S}
{Singer}, L.~P., {Price}, L.~R., {Farr}, B., {et~al.} 2014, ArXiv e-prints

\bibitem[{{Stecker} {et~al.}(2006){Stecker}, {Malkan}, \&
  {Scully}}]{Stecker+06ebl}
{Stecker}, F.~W., {Malkan}, M.~A., \& {Scully}, S.~T. 2006, \apj, 648, 774

\bibitem[{Tanvir {et~al.}(2013)Tanvir, Levan, Fruchter, Hjorth, Hounsell,
  Wiersema, \& Tunnicliffe}]{tanvir2013kilonova}
Tanvir, N., Levan, A., Fruchter, A., {et~al.} 2013, Nature, 500, 547

\bibitem[{{The LIGO Scientific Collaboration} \& {the Virgo
  Collaboration}(2013)}]{2013arXiv1310.2314T}
{The LIGO Scientific Collaboration}, \& {the Virgo Collaboration}. 2013, ArXiv
  1310.2314

\bibitem[{{Toma} {et~al.}(2011){Toma}, {Wu}, \&
  {M{\'e}sz{\'a}ros}}]{2011MNRAS.415.1663T}
{Toma}, K., {Wu}, X.-F., \& {M{\'e}sz{\'a}ros}, P. 2011, \mnras, 415, 1663

\bibitem[{{Troja} {et~al.}(2010){Troja}, {Rosswog}, \&
  {Gehrels}}]{Troja+10precursor}
{Troja}, E., {Rosswog}, S., \& {Gehrels}, N. 2010, \apj, 723, 1711

\bibitem[{{van der Sluys} {et~al.}(2009){van der Sluys}, {Mandel}, {Raymond},
  {Kalogera}, {R{\"o}ver}, \& {Christensen}}]{2009CQGra..26t4010V}
{van der Sluys}, M., {Mandel}, I., {Raymond}, V., {et~al.} 2009, Classical and
  Quantum Gravity, 26, 204010

\bibitem[{{van der Sluys} {et~al.}(2008){van der Sluys}, {R{\"o}ver},
  {Stroeer}, {Raymond}, {Mandel}, {Christensen}, {Kalogera}, {Meyer}, \&
  {Vecchio}}]{2008ApJ...688L..61V}
{van der Sluys}, M.~V., {R{\"o}ver}, C., {Stroeer}, A., {et~al.} 2008, \apjl,
  688, L61

\bibitem[{{van Eerten} \& {MacFadyen}(2011)}]{2011ApJ...733L..37V}
{van Eerten}, H.~J., \& {MacFadyen}, A.~I. 2011, \apjl, 733, L37

\bibitem[{{Veitch} {et~al.}(2012){Veitch}, {Mandel}, {Aylott}, {Farr},
  {Raymond}, {Rodriguez}, {van der Sluys}, {Kalogera}, \&
  {Vecchio}}]{2012PhRvD..85j4045V}
{Veitch}, J., {Mandel}, I., {Aylott}, B., {et~al.} 2012, \prd, 85, 104045

\bibitem[{{Veres} \& {M{\'e}sz{\'a}ros}(2013)}]{Veres+13tev}
{Veres}, P., \& {M{\'e}sz{\'a}ros}, P. 2013, ArXiv 1312.0590

\bibitem[{{Veres} {et~al.}(2013){Veres}, {Zhang}, \&
  {M{\'e}sz{\'a}ros}}]{2013ApJ...764...94V}
{Veres}, P., {Zhang}, B.-B., \& {M{\'e}sz{\'a}ros}, P. 2013, \apj, 764, 94

\bibitem[{{Vitale} \& {Zanolin}(2011)}]{2011PhRvD..84j4020V}
{Vitale}, S., \& {Zanolin}, M. 2011, \prd, 84, 104020

\bibitem[{{Waxman} \& {Bahcall}(1997)}]{1997PhRvL..78.2292W}
{Waxman}, E., \& {Bahcall}, J. 1997, \prl, 78, 2292

\bibitem[{{Wen} \& {Chen}(2010)}]{2010PhRvD..81h2001W}
{Wen}, L., \& {Chen}, Y. 2010, \prd, 81, 082001

\bibitem[{{White} {et~al.}(2011){White}, {Daw}, \&
  {Dhillon}}]{2011CQGra..28h5016W}
{White}, D.~J., {Daw}, E.~J., \& {Dhillon}, V.~S. 2011, Classical and Quantum
  Gravity, 28, 085016

\bibitem[{{Zhang} \& {M{\'e}sz{\'a}ros}(2001{\natexlab{a}})}]{Zhang+01mag}
{Zhang}, B., \& {M{\'e}sz{\'a}ros}, P. 2001{\natexlab{a}}, \apjl, 552, L35

\bibitem[{{Zhang} \&
  {M{\'e}sz{\'a}ros}(2001{\natexlab{b}})}]{2001ApJ...559..110Z}
---. 2001{\natexlab{b}}, \apj, 559, 110

\bibitem[{{Zhang} {et~al.}(2011){Zhang}, {Zhang}, {Liang}, {Fan}, {Wu},
  {Pe'er}, {Maxham}, {Gao}, \& {Dong}}]{2011ApJ...730..141Z}
{Zhang}, B.-B., {Zhang}, B., {Liang}, E.-W., {et~al.} 2011, \apj, 730, 141

\end{thebibliography}
\end{document}